\documentclass[twocolumn,showkeys,showpacs,preprintnumbers,prd,superscriptaddress,nofootinbib]{revtex4-1}
\usepackage{graphicx}
\usepackage{epsf}
\usepackage{bm}
\usepackage{amsmath}
\usepackage{amsfonts}
\usepackage{amssymb}
\usepackage{epstopdf}
\usepackage{natbib}
\usepackage{hyperref}
\usepackage{color}
\usepackage{verbatim}
\usepackage{multirow}
\usepackage{bm}
\usepackage{hyperref}
\usepackage{float}
\usepackage{tabularx}

\definecolor{darkblue}{rgb}{0.0, 0.0, 0.55}
\definecolor{darkred}{rgb}{0.55, 0.0, 0.0}

\usepackage{hyperref}
\hypersetup{
    colorlinks=true, 
    linkcolor=darkblue,
    citecolor=darkblue,
    urlcolor=darkblue}
    
\makeatletter\let\expandableinput\@@input\makeatother

\def\Mpl{M_{\rm Pl}}

\newcolumntype{Y}{>{\centering\arraybackslash}X}
\renewcommand{\arraystretch}{1.75}

\begin{document}

\title{Challenging $\Lambda$CDM: 5$\sigma$ Evidence for a Dynamical Dark Energy Late-Time Transition}

\author{Mateus Scherer}
\email{mateus.scherer@ufrgs.br}
\affiliation{Instituto de F\'{i}sica, Universidade Federal do Rio Grande do Sul, 91501-970 Porto Alegre RS, Brazil}

\author{Miguel A. Sabogal}
\email{miguel.sabogal@ufrgs.br}
\affiliation{Instituto de F\'{i}sica, Universidade Federal do Rio Grande do Sul, 91501-970 Porto Alegre RS, Brazil}

\author{Rafael C. Nunes}
\email{rafadcnunes@gmail.com}
\affiliation{Instituto de F\'{i}sica, Universidade Federal do Rio Grande do Sul, 91501-970 Porto Alegre RS, Brazil}
\affiliation{Divisão de Astrofísica, Instituto Nacional de Pesquisas Espaciais, Avenida dos Astronautas 1758, São José dos Campos, 12227-010, São Paulo, Brazil}

\author{Antonio De Felice}
\email{antonio.defelice@yukawa.kyoto-u.ac.jp}
\affiliation{Center for Gravitational Physics and Quantum Information, Yukawa Institute for Theoretical Physics, Kyoto University, 606-8502, Kyoto, Japan}

\begin{abstract}
Recently, there has been considerable debate regarding potential evidence for the dynamical nature of dark energy (DE), particularly in light of Baryon Acoustic Oscillations (BAO) measurements released by DESI survey. In this work, we propose an agnostic test that simultaneously constrains the dark energy (DE) equation of state (EoS) and probes the possibility of a transition between the quintessence and phantom regimes, or vice versa. Our initial approach is independent of physical priors, allowing the data to determine which behavior best fits the parameters. We then consider a minimally modified gravity theory known as VCDM, into which we can map our initial approximation, placing it within a theoretically stable framework. To this end, we incorporate the most up-to-date datasets available, including BAO measurements from DESI-DR2, Type Ia Supernovae from the PantheonPlus, DESY5, and Union3 samples, as well as Cosmic Microwave Background (CMB) data from Planck. Our analysis reveals strong and statistically significant evidence for a quintessence-phantom transition across various data combinations. \textit{The strongest evidence is found for Planck+DESI+DESY5, with a significance exceeding $\sim$5$\sigma$ in favor of a quintessence-phantom transition at $z_{\dag} = 0.493^{+0.063}_{-0.081}$}. Beyond this redshift, the EoS remains within the phantom regime, while for $z < z_{\dag}$, it favors the quintessence regime. Despite this strong indication, \textit{we find that such transitions do not resolve the $H_0$ tension}.
\end{abstract}



\maketitle

\section{Introduction}

The standard cosmological model, known as the \(\Lambda\)CDM model, has been remarkably successful in explaining a wide range of cosmological observations over recent decades \cite{Planck:2018nkj,Planck:2018vyg,Planck:2019nip,Mossa:2020gjc,ACT:2020gnv,eBOSS:2020yzd}. However, with the increasing precision of modern observations, several discrepancies have emerged that challenge the \(\Lambda\)CDM framework. The most prominent of these is the so-called Hubble tension \cite{Riess:2021jrx,Kamionkowski:2022pkx,Verde:2023lmm,DiValentino:2024yew,Breuval:2024lsv,Li:2024yoe,Murakami:2023xuy,Verde:2019ivm,Knox:2019rjx,DiValentino:2020zio,DiValentino:2021izs,DiValentino:2022fjm,Said:2024pwm,Boubel:2024cqw,Freedman:2024eph,Riess:2024vfa,Scolnic:2024hbh}, which refers to the discrepancy in the measurement of the Hubble constant \(H_0\). Local measurements, including those from Cepheid variable stars and Type Ia supernovae, provide values of \(H_0\) that are significantly higher than those inferred from the Cosmic Microwave Background (CMB) data assuming the \(\Lambda\)CDM model, with the discrepancy reaching a significance greater than 5\(\sigma\) \cite{Riess:2021jrx,Breuval:2024lsv}.

In addition to the ongoing debate surrounding the Hubble tension, another key challenge in modern cosmology is the discrepancy in the parameter $S_8$. This parameter, defined as $S_8 = \sigma_8 \sqrt{\Omega_m/0.3}$, relates the amplitude of matter fluctuations to the total matter density $\Omega_m$. Early analyses of cosmic shear measurements suggested a tension exceeding $3\sigma$~\cite{KiDS:2020suj,DES:2017qwj}, but recent results from the KiDS survey indicate strong consistency with the $\Lambda$CDM model~\cite{Wright:2025xka,Stolzner:2025htz}.  Similarly, constraints on $S_8$ derived from ACT-CMB data align well with Planck-CMB predictions, further reinforcing the standard cosmological scenario~\cite{louis2025atacama}. Other observational probes also support $\Lambda$CDM at the $S_8$ level~\cite{Sailer:2024coh,Chen:2024vvk,Anbajagane:2025hlf,Garcia-Garcia:2024gzy,Sugiyama:2023fzm}. However, full-shape analyses of large-scale structure data reveal a more significant tension, at the level of at least $4.5\sigma$~\cite{Ivanov:2024xgb,Chen:2024vuf}, suggesting a potential discrepancy between early- and late-time universe constraints. Moreover, studies based on Redshift-Space Distortions (RSD) have also pointed to deviations that may indicate additional inconsistencies~\cite{Nunes:2021ipq,Kazantzidis_2018}. Further observational probes have reported evidence of possible tensions in $S_8$ as well~\cite{Karim:2024luk,Dalal:2023olq,Akarsu:2024hsu,Adil:2023jtu,Giare:2025ath}. The persistence of these discrepancies across various datasets raises important questions about the completeness of the $\Lambda$CDM model, potentially requiring modifications such as refinements in the modeling of the matter power spectrum, alternative DE scenarios, modifications to gravity, or interactions in the dark sector (see~\cite{DiValentino:2025sru} for a general review).

On the other hand, observational efforts aimed at investigating the possibility of a late-time cosmological transition have long garnered interest—both as a probe of the fundamental nature of DE and as a potential resolution to the $H_0$ tension (see, for instance, \cite{Akarsu:2023mfb,Soriano:2025gxd,Souza:2024qwd,Gomez-Valent:2024ejh, Akarsu:2024eoo,DiValentino:2019exe,Hu:2022kes,Mukherjee:2024akt,Alestas:2022xxm,Tsujikawa:2006ph,Bassett:2002qu,Martins:2018bzo,Sabogal:2025mkp,Hu:2022kes,Alestas:2021luu,Liu:2024vlt,Vagnozzi:2023nrq,Huang:2024erq,Manoharan:2025uix,Lee:2022cyh,Dwivedi:2024okk,Specogna:2025guo,Yashiki:2025loj,Liu:2025mub,Zhou:2025kws}). 
Therefore, possible transitions in the dynamics of the universe related to the nature of dark energy have long been of great interest for observational tests.

From a recent observational standpoint, the Dark Energy Spectroscopic Instrument (DESI) has played a pivotal role~\cite{DESI:2025fxa, DESI:2023ytc, DESI:2025fxa}. DESI is a state-of-the-art spectroscopic survey designed to map the large-scale structure (LSS) of the Universe in three dimensions, covering an extensive area of the sky across a wide redshift range. DESI has become the largest spectroscopic dataset of its kind to date, incorporating observations of millions of galaxies and quasars. This unprecedented dataset enables the construction of detailed 3D maps of the cosmic web, facilitates precise measurements of the Universe’s expansion history~\cite{DESI:2025zgx, DESI:2024mwx}, and provides robust constraints on the dark energy equation-of-state (EoS) parameter~\cite{DESI:2025kuo, DESI:2024aqx,Garcia-Quintero:2025qeq,Ahlen:2025kat}. Additionally, the data support a wide array of cosmological analyses~\cite{DESI:2024hhd, Ishak:2024jhs, DESI:2025ejh, DESI:2024kob, Silva:2025hxw, Shah:2025ayl, Pang:2025lvh, Paliathanasis:2025dcr, Hur:2025lqc, DESI:2025hce, Anchordoqui:2025fgz, Pan:2025psn, Jiang:2024viw, Jiang:2024xnu, Chakraborty:2025syu, Pan:2025qwy, You:2025uon, Afroz:2025iwo, Wang:2025owe, Adolf:2025zar, Dinda:2025iaq, Kumar:2025etf, Giare:2025pzu, Teixeira:2025czm,Ye:2025ulq,RoyChoudhury:2025dhe,Wolf:2025jed,Wolf:2024stt,Alvarez:2025kma,Wang:2025zri,Liu:2025mub,Specogna:2025guo,deSouza:2025rhv,Toda:2025dzd,Colgain:2025nzf,Gialamas:2024lyw,Cheng:2025lod, Luciano:2025elo, Mukherjee:2025ytj, Paliathanasis:2025nmj, Paliathanasis:2025hjw, Addazi:2025qra, Cai:2025mas,Ling:2025lmw,Lee:2025hjw,Hussain:2025nqy,VanRaamsdonk:2025wvj,An:2025vfz,Li:2025eqh,vanderWesthuizen:2025iam,Lin:2025gne,Tyagi:2025zov,Wu:2025wyk,Li:2024qso,Sabogal:2025jbo,Gialamas:2025pwv,Ozulker:2025ehg,An:2025vfz,Bayat:2025xfr,Hussain:2025vbo}.

This paper investigates the possibility of late-time transitions in DE EoS. We propose an agnostic, data-driven test to probe whether DE undergoes a transition at low redshifts, characterized by a critical redshift \( z_{\dag} \). A key feature of our approach is that it does not rely on prior assumptions about the dynamics or physical nature of the EoS, thereby enabling purely observational constraints on whether the transition behaves like quintessence or phantom energy. We consider both abrupt and smooth transition models, with the latter motivated by the Lagrangian of a type II minimally modified gravity theory known as VCDM \cite{DeFelice:2020eju, DeFelice:2020cpt}. In what follows, and throughout the discussion of our main results, we refer to these two approaches as $w_{\dag}$CDM and $w_{\dag}$VCDM, respectively. Both frameworks are employed in all our statistical analyses, with a mathematically equivalent smooth version developed to establish a robust theoretical and predictive structure aimed at explaining the cosmological origin of the observed behavior. As we will show, our analysis presents—for the first time to the best of our knowledge—evidence for a transition from quintessence to phantom behavior at \( z_{\dag} \sim 0.5 \), with a significance exceeding \(\sim 5\sigma\).

This work is structured as follows: Sec.~\ref{model} presents the theoretical model and its main physical characteristics. Sec.~\ref{data} describes the methodology and datasets employed in our analysis. The results are presented in Sec.~\ref{results}, and our final conclusions are summarized in Sec.~\ref{conclusions}.

\section{Agnostic test for probing late-time transitions}
\label{model}
We consider a homogeneous and isotropic model of the Universe, described by the Friedmann-Lema\^itre-Robertson-Walker (FLRW) metric, which serves as the foundation for modern cosmological models. The line element is given by:

\begin{equation}
    ds^2 = -dt^2 + a^2(t) \left( \frac{dr^2}{1 - \kappa r^2} + r^2 d\theta^2 + r^2 \sin^2 \theta \, d\phi^2 \right),
\end{equation}
where \( a(t) \) is the scale factor governing the expansion of the Universe, and \( \kappa \) is the curvature parameter that determines the spatial geometry. Specifically, \( \kappa = 0 \) corresponds to a flat Universe, \( \kappa > 0 \) to a closed Universe with spherical spatial geometry, and \( \kappa < 0 \) to an open Universe with hyperbolic spatial curvature. 

Since recent observations strongly support a spatially flat Universe, we adopt \( \kappa = 0 \) in our analysis. Under this assumption, the expansion rate of the Universe follows the Friedmann equation:

\begin{equation}
    \mathcal{H}^2 = \frac{8\pi G}{3} a^2 \sum_i \rho_i,
\end{equation}
where \( \mathcal{H} \) is the Hubble parameter in conformal time units, \( G \) is the gravitational constant, and the summation runs over all components \( i \), including radiation (photons and neutrinos), baryons, cold dark matter (DM), and dark energy (DE). The total energy density consists of these four components, and we denote \( \rho_i \) as the energy density of the \( i \)-th component, where \( i \) runs over radiation (\( r \)), baryons (\( b \)), cold dark matter (\( c \)), and dark energy (\( x \)). Similarly, we denote \( p_i \) as the pressure of the \( i \)-th component.

The evolution of the DE energy density \( \rho_x(z) \) depends on its equation of state (EoS). Assuming a constant \( w \), the energy density evolves as:
\begin{equation}
    \rho_x(z) = \rho_{x,0} (1 + z)^{3(1+w)},
    \label{eq:rho_x}
\end{equation}  
where \( \rho_{x,0} \) is the present-day DE energy density. For \( w = -1 \), corresponding to a cosmological constant, \( \rho_x(z) \) remains constant. However, deviations from this value indicate a dynamic behavior of DE. 

In this work, we propose a general method—independent of priors and initial theoretical assumptions—to test whether a transition in the dynamic nature of DE occurs at late times (characterized by a critical redshift \( z_{\dag} \)) and to determine its direction (quintessence or phantom). Specifically, we propose that the EoS of DE undergoes an abrupt change, controlled by the parameter \( \Delta \) at a critical scale factor \( a_{\dag} \), given by:  
\begin{equation}
    \label{w_new}
    w(a) = -1 + \Delta \operatorname{sgn}\left[a_{\dagger} - a\right] \, .
\end{equation}

\begin{figure*}[htpb!]
    \centering
    \includegraphics[width=0.46\linewidth]{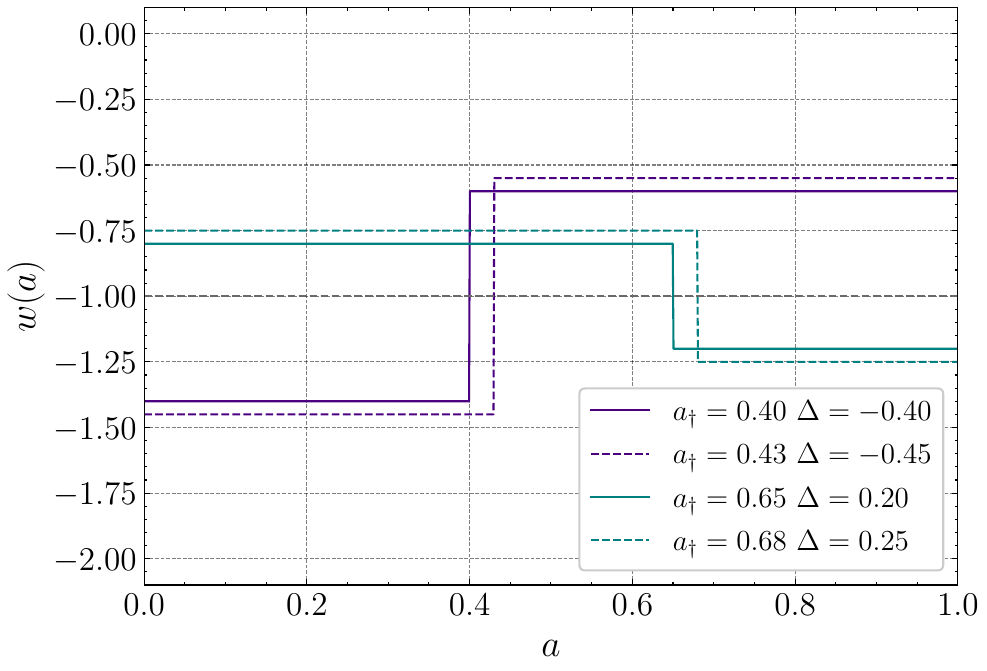} \,\,\,\,\,
    \includegraphics[width=0.46\linewidth]{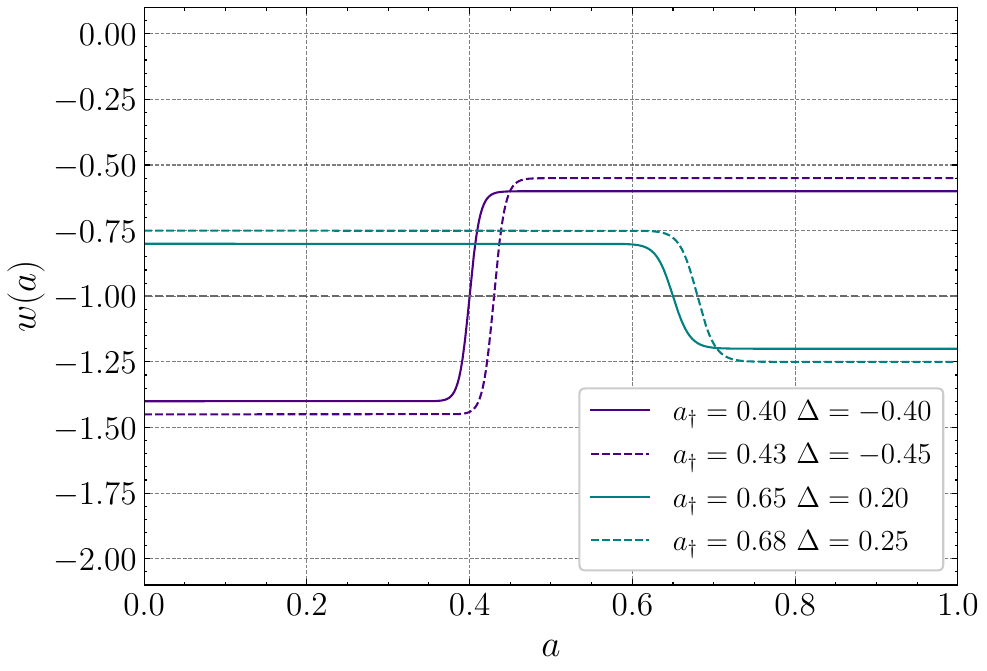}
        \caption{Left panel: Evolution of \( w(a) \) illustrating the abrupt transition for arbitrary values of \( a_{\dag} \) and \( \Delta \). Right panel: Smooth transition model Eq.~\eqref{EoS_smooth} for arbitrary \( a_{\dag} \) and \( \Delta \), with \( \zeta = 10^{1.5} \).}
    \label{fig:w_abrupt}
\end{figure*}


To implement this approach systematically, we follow these steps:

\begin{itemize}
  \item \textbf{Step 1:} Both the timing of the transition described by Eq.~\eqref{w_new} and the sign of \( \Delta \) are left entirely free to be determined by the data through the Markov Chain Monte Carlo (MCMC) analysis. In other words, no specific quintessence-to-phantom or phantom-to-quintessence transition is imposed \textit{a priori}. The preference for a particular behavior emerges solely from the evolution of the Markov chains, based on the data's statistical support. Thus, our analysis is free from any assumed dynamical behavior; the selection of a specific dynamical path arises strictly from the best fit to the observational data.

  \item \textbf{Step 2:} Once the initial value of \( \Delta \) is obtained, its sign is reversed at the critical scale factor \( a_{\dag} \) at each point in the MCMC evolution. This procedure is applied throughout all MCMC steps until the chain reaches excellent statistical convergence.
  
  \item \textbf{Step 3:} The magnitude of the variation remains constant before and after the transition. In other words, the absolute value of \( \Delta \) does not change; only its sign is reversed at the transition. A simpler strategy may also be considered, in line with conventional parametric models commonly studied in the literature. Sufficient conditions to ensure the robustness of the various analyses will be examined.

\end{itemize}


Figure~\ref{fig:w_abrupt} (left panel) shows the evolution of \( w(a) \) undergoing an abrupt transition, illustrating the model’s predictive behavior for different values of \( \Delta \) and \( a_{\dag} \). This functional form enforces a piecewise behavior in which \( w(a) \) shifts from \( w = -1 + \Delta \) for \( a < a_{\dag} \) to \( w = -1 - \Delta \) for \( a > a_{\dag} \). Here, \( \Delta \) quantifies deviations from the cosmological constant, while \( a_{\dag} \) denotes the transition epoch. By treating \( \Delta \) and \( a_{\dag} \) as free parameters—without imposing specific assumptions on the dynamics of dark energy—this framework provides an agnostic test for potential late-time transitions in the dark energy sector.

From the continuity equation, the DE energy density before the transition (\( a < a_{\dag} \)) follows:
\begin{equation}
    \rho_{\rm x} =\rho_{\rm x, 0}\, a^{-3\Delta} \, a_{\dag}^{6\Delta},
    \label{rho_x_new}
\end{equation}
where the standard solution is recovered for \( a_{\dag} = 1 \), implying no transition in the EoS.

Beyond altering the evolution of the cosmic background, a DE equation of state \( w \neq -1 \) also impacts the growth of perturbations. To analyze these effects, we work in the synchronous gauge, where the perturbed FLRW metric is given by  
\begin{equation}
ds^2 = a^2(\eta) \left[ -d\eta^2 + (\delta_{ij} + h_{ij}) dx^i dx^j \right],
\label{eq:perturbed_metric}
\end{equation}  
where \( a(\eta) \) is the scale factor, \( \eta \) represents conformal time, and \( h \equiv h_{ii} \) is the trace of the synchronous gauge metric perturbation \( h_{ij} \). Transforming to Fourier space, we define the DE density contrast as \( \delta_x \) and the velocity divergence as \( \theta_x \). Their evolution is governed by the following equations:
\begin{equation}
\begin{aligned}
\dot{\delta}_x &= -(1+w) \left(\theta + \frac{\dot{h}}{2} \right) 
- 3(\hat{c}_s^2 - w) \mathcal{H} \delta_x  \\ 
&\quad - 9(1+w)(\hat{c}_s^2 - c_a^2) \mathcal{H}^2 \frac{\theta_x}{k^2},  \\  
\dot{\theta}_x &= -(1-3\hat{c}_s^2) \mathcal{H} \theta_x 
+ \frac{\hat{c}_s^2 k^2}{1+w} \delta_x - k^2 \sigma_x.
\end{aligned}
\label{fortheta}
\end{equation}

These equations are fully general, assuming a non-interacting fluid while allowing for the presence of shear stress \( \sigma_x \), a non-adiabatic sound speed, and a time-dependent equation of state parameter \( w \). From this point forward, we assume a shear-free fluid and adopt the parametrization of \( w \) given by Eq.~(\ref{w_new}). Within the perspectives discussed above, we adopt the PPF approximation \cite{Fang:2008sn} in all subsequent analyses.

\section{Embedding the $w$-Switch model within the VCDM}

The idea of a switch for the function $w(a)$ looks interesting, but from a theoretical point of view, one may wonder how to implement it so that the background evolution remains stable at all times. For instance, for a perfect fluid with pressure $P$ and equation of state $P=P(\rho,s)$, $s$ being the entropy per particle, we know that the perturbations propagate with a speed $c_s^2=\left(\frac{\partial P}{\partial\rho}\right)_{\! s}$, see e.g.~\cite{Misner:1973prb,DeFelice:2009bx}. On a homogeneous and isotropic background, this quantity would become $c_s^2= P'/\rho'=w-\frac13\,w'/(1+w)$, where a prime denotes differentiation with respect to the e-fold variable $N=\ln(a/a_0)$. Either before or after the sign-transition for $w$, since $w'\approx0$, the value of $c_s^2$ would become negative, leading to an instability in the cosmological perturbations which would spoil the background dynamics in a time scale much shorter than the Hubble time. This shows that embedding this $w$-switch into a perfect fluid action is impossible.

However, there are theoretical models capable of embedding this $w$-switch without spoiling the results we want to present. In particular, we will show here the possibility of embedding the $w$-switch model into VCDM \cite{DeFelice:2020eju}. The theory of VCDM represents a model and framework that allows a large variety of time-dependent dark energy dynamics without introducing any new propagating gravity/matter degree of freedom. Doing so immediately implies that one does not need to worry about unstable or ghost-like degrees of freedom, just because they do not exist in the theory by construction. Several theoretical studies have been performed to understand this model, both in the context of compact objects and cosmology \cite{DeFelice:2021xps, DeFelice:2022riv, Jalali:2023wqh, DeFelice:2022uxv, DeFelice:2020cpt}, and other dark energy dynamical models have already been embedded in this framework \cite{Akarsu:2024qsi, Akarsu:2024eoo, DiValentino:2025sru}. Since the theoretical work has already been implemented elsewhere, here we just write down the two differences in the dynamics of background/perturbations the model presents concerning $\Lambda$CDM. First, consider the background evolution: the $w$-transition must be smooth. To simplify the calculations, we adopt the following smooth functional form for \( w(N) \):
\begin{equation}
\label{EoS_smooth}
    w(N) = -1 + \Delta \tanh\left[\zeta(N_\dagger - N)\right]\,,
\end{equation}
where \( \zeta \) is the parameter that controls the rapidity and smoothness of the transition, and \( N \) is the e-fold time variable defined by \( N = \ln(a/a_0) \), with \( N \leq 0 \). A larger \( \zeta \) corresponds to a sharper transition, while a smaller \( \zeta \) yields a more gradual change in \( w(a) \). Figure~\ref{fig:w_abrupt} (right panel) shows the evolution of \( w(a) \) under this approximation for the equation of state. With this choice, an analytical expression for \( \rho_x \) can be obtained, namely

\begin{align}
    \varrho_x&=C\,(\cosh[\zeta(N-N_\dagger)])^{3\Delta/\zeta}\nonumber\\
    &=C\left[\frac{1}{2}\left(\frac{a}{a_\dagger}\right)^\zeta+\frac{1}{2}\left(\frac{a_\dagger}{a}\right)^\zeta\right]^{3\Delta/\zeta},
\end{align}
where $C$ is a constant. Now the modified Friedmann equation reads
\begin{equation}
    H(a)^2=\sum_I\varrho_I(a) + \varrho_x(a)\,,
\end{equation}
where $H$ is the Hubble factor ($H=\dot{a}/a^2$ and a dot stands for derivative with respect to the conformal time), the index/label $I$ runs over all the standard model matter components, and we have normalized for each matter component $\varrho=\rho/(3\Mpl^2)$. The constant $C$ is determined by imposing $H_0^2\bigl(1-\sum_I\Omega_{I0}\bigr)=\varrho_x(a=a_0)$. The pressure is given, by construction, by the expression $p_x=w\, \varrho_x$, here normalized, like the energy density, as $p_x=P_x/(3\Mpl^2)$.

The equations of motion for the cosmological perturbations, written here in the Newtonian gauge, are identical to those in the $\Lambda$CDM model (noting that VCDM does not introduce any additional perturbation modes), except for the momentum constraint, which can be expressed—without approximations—as
\begin{equation}
\dot{\Phi}+aH\Psi  = 
\frac{3\,[k^{2}-3a^{2}(\dot{H}/a)]\sum_{I}(\varrho_{I}+p_{I})\,\theta_{I}}{k^{2}\,[2k^{2}/a^{2}+9\sum_{K}(\varrho_{K}+p_{K})]}\,, \label{eqn:dotphi}
\end{equation}
where $\Phi$ and $\Psi$ are the two Bardeen potentials, $\theta$ is the scalar perturbation of the fluid velocity, and, once more, the indices $I$ and $K$ run over all the standard model matter species (including the contribution from dark matter) except for the $x$-component. The difference here stands in the presence of the conformal time derivative of $H$, which differs from the expression valid in $\Lambda$CDM, i.e.\ $\dot{H}/a\neq -\tfrac32\sum_{I}(\varrho_{I}+p_{I})$.\footnote{The $x$-component does satisfy the continuity equation, $\dot{\varrho}_x/a=-3H(\varrho_x + p_x)$, but it is not a cosmological constant, $\varrho_x+p_x\neq0$.}  Having stated the theory, we can proceed with the data analysis.

In what follows, we explore the observational constraints for both frameworks defined above, namely the $w_{\dag}$CDM and $w_{\dag}$VCDM models.

\section{Datasets and methodology}
\label{data}
To rigorously test our proposed theoretical framework, we implemented it within the Boltzmann solver \texttt{CLASS}~\cite{Blas:2011rf} and performed Markov Chain Monte Carlo (MCMC) analyses using the publicly available sampler \texttt{MontePython}~\cite{Brinckmann:2018cvx, Audren:2012wb}. Convergence was ensured in all runs by imposing the Gelman-Rubin criterion~\cite{Gelman_1992} with $R-1 \leq 10^{-2}$. We adopted flat priors on the cosmological parameter set \{$\Omega_{\rm b} h^2$, $\Omega_{\rm c} h^2$, $\tau_{\rm reio}$, $100\theta_{\mathrm{s}}$, $\log(10^{10} A_{\mathrm{s}})$, $n_{\mathrm{s}}$, $\Delta$, $z_{\dag}$\}, where the first six are standard $\Lambda$CDM parameters. Specifically, these include the present-day physical densities of baryons ($\omega_b = \Omega_{\rm b} h^2$) and cold dark matter ($\omega_c = \Omega_{\rm c} h^2$), the reionization optical depth ($\tau_{\rm reio}$), the angular scale of the sound horizon at recombination ($\theta_{\mathrm{s}}$), the amplitude of primordial scalar perturbations ($A_{\mathrm{s}}$), and the spectral index of scalar fluctuations ($n_{\mathrm{s}}$). The additional parameters $\Delta$ and $z_{\dag}$ characterize the transition in the dark energy equation of state. The prior ranges were set as follows: $\omega_b \in [0.0, 1.0]$, $\omega_{\text{cdm}} \in [0.0, 1.0]$, $100 \theta_s \in [0.5, 2.0]$, $\ln(10^{10} A_s) \in [1.0, 5.0]$, $n_s \in [0.1, 2.0]$, $\tau_{\text{reio}} \in [0.004, 0.8]$, $\Delta \in [-5, 5]$, and $z_{\dag} \in [0, 10]$. Throughout all analyses conducted for the VCDM scenario, the parameter $\zeta$ is set to $10^{1.5}$. For statistical analysis and visualization of the MCMC chains, we employed the Python package \texttt{GetDist}\footnote{\url{https://github.com/cmbant/getdist}}, which we used to derive 1D posterior distributions and 2D confidence contours. \footnote{Throughout this study, we expanded the prior ranges by up to an order of magnitude beyond the values reported here and those commonly used in standard analyses. We observed no significant impact on the results, confirming the robustness of our conclusions to the choice of prior ranges.}

The datasets used in the analyses are described below.
\begin{itemize}

\item \textit{Cosmic Microwave Background} (\textbf{CMB}): We utilize temperature and polarization anisotropy measurements of the CMB power spectra from the Planck satellite~\cite{Planck:2018vyg}, along with their cross-spectra from the Planck 2018 legacy data release. Specifically, we employ the high-$\ell$ \texttt{Plik} likelihood for TT (in the multipole range $30 \leq \ell \leq 2508$), TE, and EE ($30 \leq \ell \leq 1996$), as well as the low-$\ell$ TT-only ($2 \leq \ell \leq 29$) likelihood and the low-$\ell$ EE-only ($2 \leq \ell \leq 29$) \texttt{SimAll} likelihood~\cite{Planck:2019nip}. Additionally, we include CMB lensing measurements, which are reconstructed from the temperature 4-point correlation function~\cite{Planck:2018lbu}. We refer to this dataset as \texttt{CMB}.

\item \textit{Baryon Acoustic Oscillations} (\textbf{DESI DR2}): We consider BAO measurements from DESI’s second data release, which includes observations of galaxies and quasars~\cite{DESI:2025zgx}, as well as Lyman-$\alpha$ tracers~\cite{DESI:2025zpo}. These measurements, detailed in Table IV of Ref.~\cite{DESI:2025zgx}, cover both isotropic and anisotropic BAO constraints over $0.295 \leq z \leq 2.330$, divided into nine redshift bins. The BAO constraints are expressed in terms of the transverse comoving distance $D_{\mathrm{M}}/r_{\mathrm{d}}$, the Hubble horizon $D_{\mathrm{H}}/r_{\mathrm{d}}$, and the angle-averaged distance $D_{\mathrm{V}}/r_{\mathrm{d}}$, all normalized to the comoving sound horizon at the drag epoch, $r_{\mathrm{d}}$. We also incorporate the correlation structure of these measurements through the cross-correlation coefficients: $r_{V,M/H}$, capturing the correlation between $D_{\mathrm{V}}/r_{\mathrm{d}}$ and $D_{\mathrm{M}}/D_{\mathrm{H}}$, and $r_{M,H}$, which describes the correlation between $D_{\mathrm{M}}/r_{\mathrm{d}}$ and $D_{\mathrm{H}}/r_{\mathrm{d}}$. This dataset is referred to as \texttt{DR2}.

\item \textit{Type Ia Supernovae} (\textbf{SN Ia}): Type Ia supernovae act as standardizable candles, providing a crucial method for measuring the universe's expansion history. Historically, SN Ia played a pivotal role in the discovery of the accelerating expansion of the universe~\cite{SupernovaSearchTeam:1998fmf, perlmutter1999measurements}, building upon earlier, more complex arguments supporting $\Lambda$-dominated models from large-scale structure observations. In this work, we will use the following recent samples:
\begin{enumerate}
    \item [(i)] \textbf{PantheonPlus and PantheonPlus\&SH0ES}: We integrated the latest distance modulus measurements from SN Ia in the PantheonPlus sample~\cite{pantheonplus}, which includes 1701 light curves from 1550 distinct SN Ia events, spanning a redshift range of $0.01$ to $2.26$. We designate this dataset as \texttt{PP}. Additionally, we consider a version of this sample that uses the latest SH0ES Cepheid host distance anchors~\cite{Riess:2021jrx} to calibrate the absolute magnitude of SN Ia, rather than centering a prior on the $H_{0}$ value from SH0ES. This approach allows for more robust results, and this version of the dataset is referred to as \texttt{PPS}. 

    \item [(ii)] \textbf{Union 3.0}: The Union 3.0 compilation, consisting of 2087 SN Ia within the range $0.001 < z < 2.260$, was presented in~\cite{Rubin:2023ovl}. Notably, 1363 of these SN Ia overlap with the PantheonPlus sample. This dataset features a distinct treatment of systematic errors and uncertainties, employing Bayesian Hierarchical Modeling. We refer to this dataset as \texttt{Union3}.

    \item [(iii)] \textbf{DESY5}: As part of their Year 5 data release, the Dark Energy Survey (DES) recently published results from a new, homogeneously selected sample of 1635 photometrically-classified SN Ia with redshifts spanning $0.1 < z < 1.3$~\cite{DES:2024tys}. This sample is complemented by 194 low-redshift SN Ia (shared with the PantheonPlus sample) in the range $0.025 < z < 0.1$. We refer to this dataset as \texttt{DESY5}.
\end{enumerate}
\end{itemize}

\subsection{Statistical tests and significance}
\label{Statistical_tests}
When discussing the statistical significance of our model compared to the standard $\Lambda$CDM scenario, we perform a chi-square difference test. Beyond this simple comparison, since the models in observational tests in this work have different numbers of free parameters, the standard model has $N_1 = 6$ free parameters. In contrast, our extended model introduces $N_2 = 8$, resulting in a difference in degrees of freedom $\Delta N = N_2 - N_1 = 2$. The difference in chi-square values, given by  
\begin{equation}
\Delta\chi^2 = \chi^2_{\text{our model}} - \chi^2_{\Lambda \text{CDM}} ,
\end{equation}
follows a chi-square distribution with $\Delta N$ degrees of freedom, under the null hypothesis that the additional parameters do not improve the fit.  

To determine the statistical significance, we compute the \( p \)-value, which is defined as the probability of obtaining a chi-square value greater than \(|\Delta\chi^2|\). Then, to express this significance in terms of standard deviations, we use the inverse cumulative distribution function of the standard normal distribution, \( \Phi^{-1} \). The corresponding significance in terms of standard deviations, \( \sigma \), is calculated as:

\begin{equation}
\sigma = \Phi^{-1} \left( 1 - \frac{p}{2} \right),
\end{equation}
where the factor \( \frac{p}{2} \) accounts for the two-tailed significance test.

To further assess the level of agreement (or disagreement) between the models and their association with each analyzed dataset, we also perform a statistical comparison of the model with the $\Lambda$CDM scenario using the well-known Akaike Information Criterion (AIC) \cite{Akaike:1974vps}, defined as:  
\begin{equation}
    \text{AIC} \equiv -2 \ln L_{\text{max}} + 2N,
\end{equation}
where $L_{\text{max}}$ is the maximum likelihood of the model, and $N$ is the total number of free parameters. A lower AIC value indicates a better model, balancing both goodness of fit and model complexity. Specifically, the AIC incorporates a penalty term for the number of parameters, discouraging the inclusion of unnecessary parameters that could lead to overfitting.


In addition to the AIC, we also consider the Bayesian Evidence, which provides a more rigorous assessment by computing the Bayes factor, quantifying the support for one model relative to another. Given a dataset $\mathbf{x}$ and two competing models, $\mathcal{M}_i$ and $\mathcal{M}_j$, with parameter sets $\boldsymbol{\theta}_i$ and $\boldsymbol{\theta}_j$, respectively, the Bayes factor $\mathcal{B}_{ij}$ (assuming equal prior probabilities for the models) is defined as:

\begin{equation}
\mathcal{B}_{ij} = \frac{p(\mathcal{M}_i|\mathbf{x})}{p(\mathcal{M}_j|\mathbf{x})} = \frac{\displaystyle \int d \boldsymbol{\theta}_i \, \pi\left(\boldsymbol{\theta}_i | \mathcal{M}_i\right) \mathcal{L}\left(\mathbf{x} | \boldsymbol{\theta}_i, \mathcal{M}_i\right)}{\displaystyle\int d \boldsymbol{\theta}_j \, \pi\left(\boldsymbol{\theta}_j | \mathcal{M}_j\right) \mathcal{L}\left(\mathbf{x} | \boldsymbol{\theta}_j, \mathcal{M}_j\right)},
\label{BayesFactor}
\end{equation}
where $p(\mathcal{M}_i|\mathbf{x})$ is the Bayesian evidence for model $\mathcal{M}_i$, $\pi(\boldsymbol{\theta}_i|\mathcal{M}_i)$ denotes the prior distribution, and $\mathcal{L}(\mathbf{x}|\boldsymbol{\theta}_i, \mathcal{M}_i)$ is the likelihood of the data given the parameters. A Bayes factor $\mathcal{B}_{ij} > 1$ favors model $\mathcal{M}_i$ (e.g., $\Lambda$CDM) over model $\mathcal{M}_j$ (e.g., DDE), even if the latter has a better fit, due to penalization for model complexity and prior volume.

To interpret the strength of evidence, we adopt the Jeffreys–Kass–Raftery scale~\cite{Kass:1995loi}:
\begin{itemize}
    \item $0 \leq |\ln \mathcal{B}_{ij}| < 1$: Weak evidence.
    \item $1 \leq |\ln \mathcal{B}_{ij}| < 3$: Positive evidence.
    \item $3 \leq |\ln \mathcal{B}_{ij}| < 5$: Strong evidence.
    \item $|\ln \mathcal{B}_{ij}| \geq 5$: Very strong evidence.
\end{itemize}

To summarize, a negative value of $\ln \mathcal{B}_{ij}$ suggests a preference for the DDE model over $\Lambda$CDM, whereas a positive value favors $\Lambda$CDM. For computing Bayes factors, we use the publicly available \texttt{MCEvidence} package~\cite{Heavens:2017hkr,Heavens:2017afc}.\footnote{Available at \url{https://github.com/yabebalFantaye/MCEvidence}.}

\section{Results and Discussions}
\label{results}
In the discussions that follow, we will discuss and present our main results. We distinguish between the results obtained with and without the inclusion of DESI DR2 data. In each case, we consider two scenarios: the \textbf{$\bm{w_\dagger}$CDM} model and its generalization, the \textcolor{blue}{\textbf{$\bm{w_{\dagger}}$VCDM}} model. For both models, we show the results for the key parameter of interest, primarily the deviation parameter $\Delta$ and its
correlations with other cosmological parameters, such as $H_{0}$ and $S_{8}$ related to cosmic tensions. Additionally, we will present a detailed discussion about the statistical significance of our results relative to $\Lambda$CDM. Despite the structural generalization introduced in $w_{\dagger}$VCDM, we find that its results remain broadly consistent with those of $w_{\dagger}$CDM. Therefore, given the strong similarity between the two cases, we will primarily refer to the $w_{\dagger}$CDM scenario in the following discussions, as the statistical conclusions are analogous for both models.

\begin{table}[htpb!]
\begin{center}
\caption{Marginalized constraints, along with the mean values at 68\% CL, for both the free and some derived parameters of the models considered in this work, based on the CMB dataset and its combinations with PP\&SH0ES. In the last rows, we present \(\Delta \chi^2_{\text{min}} = \chi^2_{\text{min (our model)}} - \chi^2_{\text{min ($\Lambda$CDM)}}\) and \(\Delta \text{AIC} = \text{AIC}_{\text{our model}} - \text{AIC}_{\text{$\Lambda$CDM}}\). Negative values for \(\Delta \chi^2_{\text{min}}\) and \(\Delta \text{AIC}\) indicate a better fit of the model compared to the \(\Lambda\)CDM model.}
\label{tab:planck_nobao}
\renewcommand{\arraystretch}{1.0}
\resizebox{\columnwidth}{!}{
\begin{tabular}{l||cccc||ccc} 
\hline
\textbf{Dataset} & \textbf{Planck} & \textbf{Planck+PP\&SH0ES} \\
\hline
\textbf{Model} & \textbf{$w_\dagger$CDM} & \textbf{$w_\dagger$CDM} \\
& \textcolor{blue}{\textbf{$w_{\dag}$VCDM}} & \textcolor{blue}{\textbf{$w_{\dag}$VCDM}} \\
\hline\hline
{$10^{2}\omega{}_{b }$} & $2.241\pm 0.016$ & $2.253\pm 0.014$ \\
& \textcolor{blue}{$2.246\pm 0.015$} & \textcolor{blue}{$2.255\pm 0.014$} \\[0.1cm]

{$\omega{}_{cdm }$} & $0.1196\pm 0.0013$ & $0.1185\pm 0.0012$ \\
& \textcolor{blue}{$0.1190\pm 0.0012$} & \textcolor{blue}{$0.1183\pm 0.0012$} \\[0.1cm]

$100\theta{}_{s }$ & $1.04189\pm 0.00030$ & $1.04205\pm 0.00030$ \\
& \textcolor{blue}{$1.04195\pm 0.00028$} & \textcolor{blue}{$1.04209\pm 0.00029$} \\[0.1cm]

$\ln10^{10}A_{s }$ & $3.042\pm 0.015$ & $3.050\pm 0.015$ \\
& \textcolor{blue}{$3.038\pm 0.014$} & \textcolor{blue}{$3.048\pm 0.015$} \\[0.1cm]

$n_{s}$ & $0.9662\pm 0.0043$ & $0.9688\pm 0.0042$ \\
& \textcolor{blue}{$0.9677\pm 0.0039$} & \textcolor{blue}{$0.9694\pm 0.0041$} \\[0.1cm]

$\tau{}_{\text{reio}}$ & $0.0538\pm 0.0073$ & $0.0582^{+0.0073}_{-0.0083}$ \\
& \textcolor{blue}{$0.0527\pm 0.0072$} & \textcolor{blue}{$0.0575^{+0.0069}_{-0.0083}$} \\[0.1cm]

$\Delta$ & $-0.29^{+0.24}_{-0.71}$ & $-0.174\pm 0.054$ \\
& \textcolor{blue}{$-0.68^{+0.15}_{-0.28}$} & \textcolor{blue}{$-0.187\pm 0.051$} \\[0.1cm]

$z_{\dagger}$ & $< 0.626$ & $0.167^{+0.036}_{-0.045}$ \\
& \textcolor{blue}{$< 0.651$} & \textcolor{blue}{$0.174^{+0.032}_{-0.044}$} \\

\hline
$\mathrm{H}_0 \, [\mathrm{km/s/Mpc}]$ & $75.0^{+7.3}_{-13}$ & $69.93\pm 0.66$ \\
& \textcolor{blue}{$79.6^{+7.8}_{-9.0}$} & \textcolor{blue}{$70.09\pm 0.66$} \\[0.1cm]

$\Omega_{\rm m}$ & $0.267^{+0.063}_{-0.079}$ & $0.2898\pm 0.0065$ \\
& \textcolor{blue}{$0.231^{+0.043}_{-0.053}$} & \textcolor{blue}{$0.2882\pm 0.0064$} \\[0.1cm]

$\mathrm{S}_{8}$ & $0.796^{+0.041}_{-0.030}$ & $0.801\pm 0.012$ \\
& \textcolor{blue}{$0.778^{+0.031}_{-0.027}$} & \textcolor{blue}{$0.799\pm 0.012$} \\

\hline
$\Delta \chi^2_{\text{min}}$ & $-3.90$ & $-12.78$ \\
& \textcolor{blue}{$-3.62$} & \textcolor{blue}{$-13.20$} \\[0.1cm]

$\Delta{\text{AIC}}$ & $\quad 0.10$ & $-8.78$ \\
& \textcolor{blue}{$\quad 0.38$} & \textcolor{blue}{$-9.20$} \\

\hline
\hline
\end{tabular}}
\end{center}
\end{table}

In Table \ref{tab:planck_nobao}, we first analyze the constraints imposed by CMB data alone and then explore its combination with the PP\&SH0ES sample. Our findings show that CMB data by itself lacks sufficient sensitivity to constrain the extended parameter space of our models. In particular, it only provides an upper bound on the transition redshift, yielding \( z_{\dag} < 0.626 \) at 68\% CL. Likewise, the deviation parameter \( \Delta \) remains weakly constrained, although there is a $\sim$1$\sigma$ tendency toward deviations favoring $\Delta < 0$ in both scenarios. The presence of new degrees of freedom introduces degeneracies that reduce the constraining power of CMB data, especially in the estimation of \( H_0 \) and \( \Omega_m \). Although the posterior distributions of these parameters shift from their \( \Lambda \)CDM values, the associated uncertainties remain large. Consequently, the Planck-only dataset does not impose strong restrictions on the new physics explored by our frameworks.

\begin{figure*}[htpb!]
    \centering
    \includegraphics[width=0.46\textwidth]{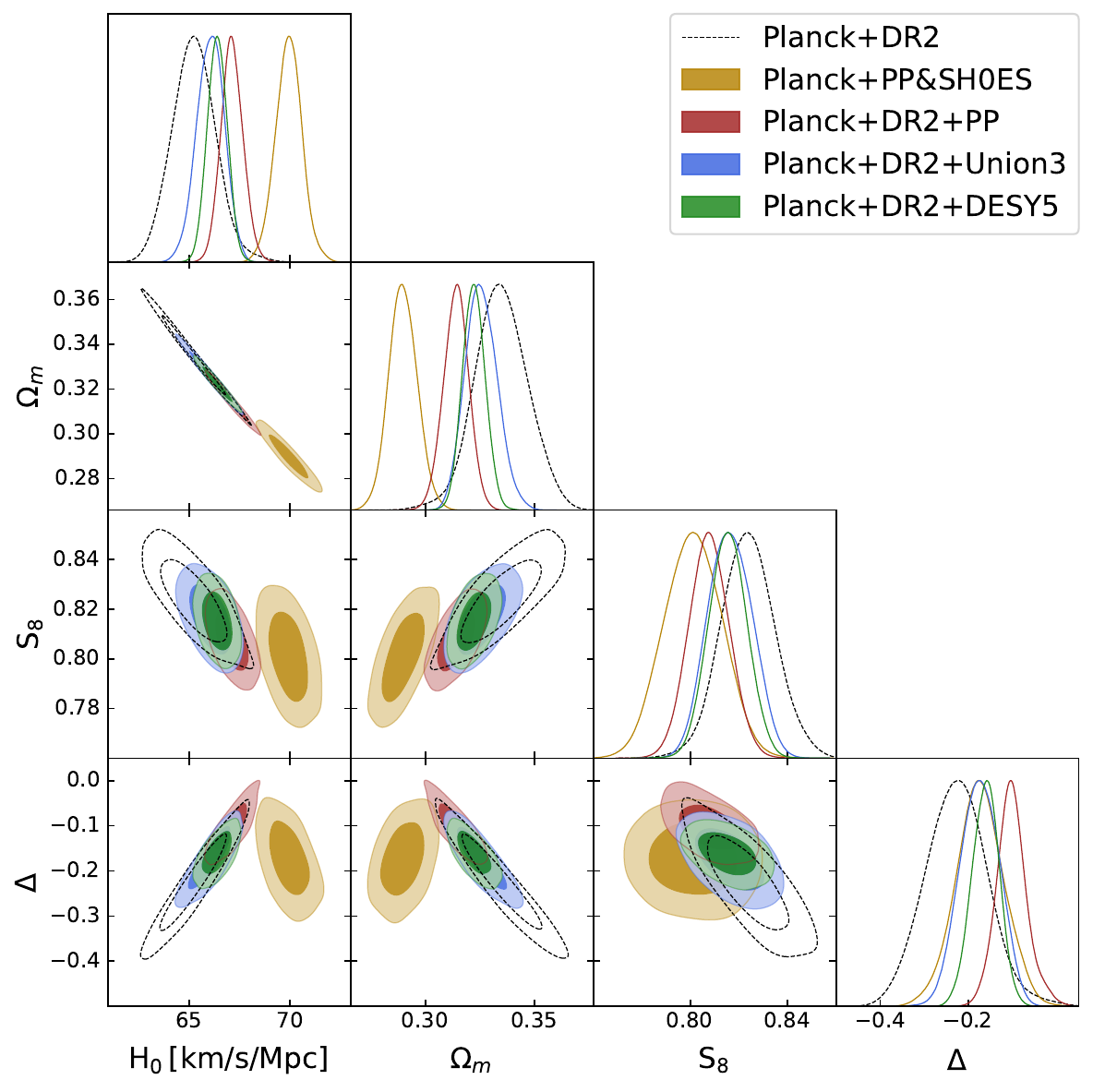}\,\,\,\,
    \includegraphics[width=0.46\textwidth]{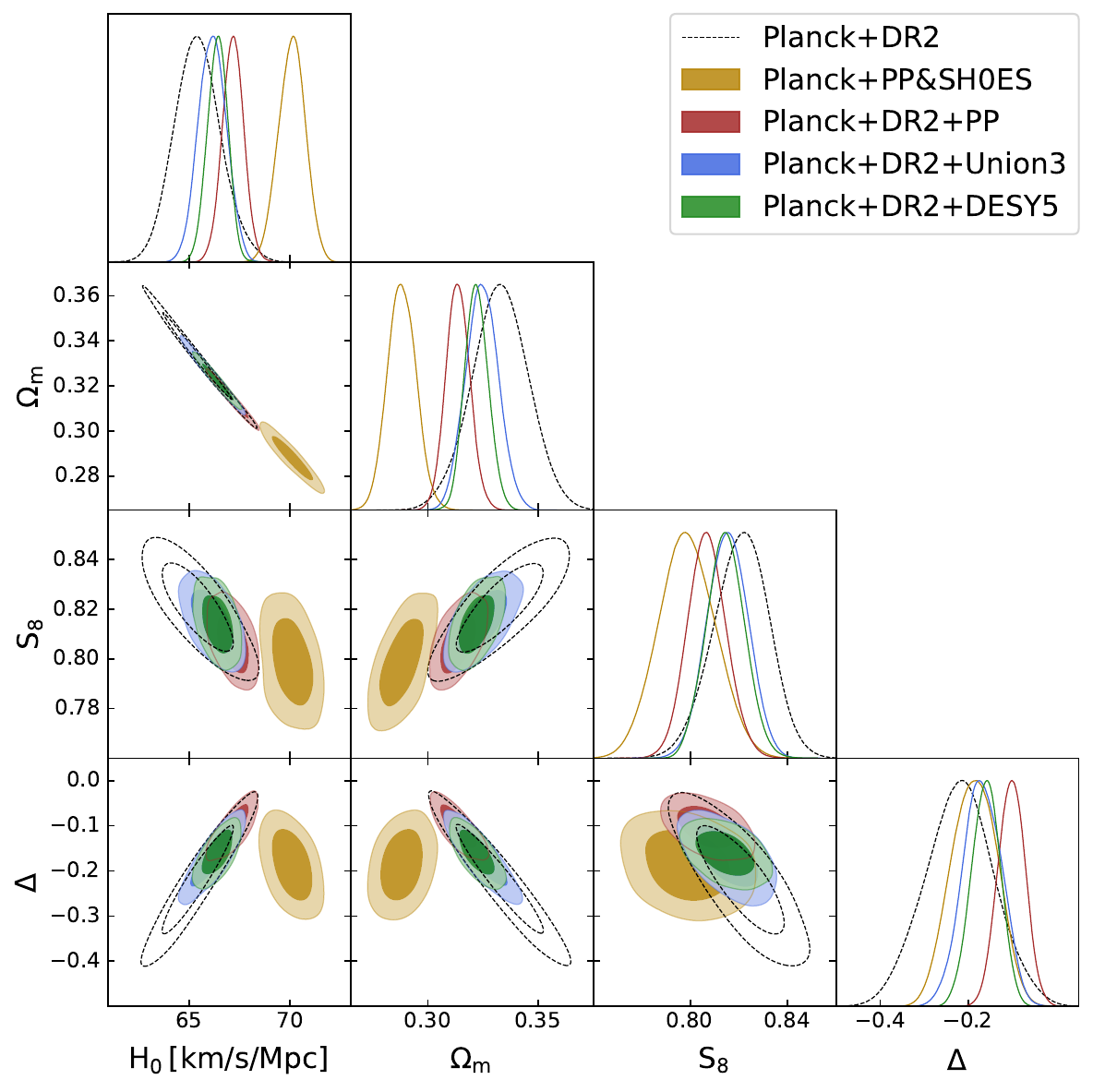}
    \caption{Left panel: Marginalized one-dimensional posterior distributions and contours (68\% and 95\% CL) for the parameters \(\Delta\), \(H_0\), \(S_8\), and \(\Omega_m\) in the \(w_{\dagger}\)CDM model, based on the CMB dataset and its combinations with PP\&SH0ES, DESI, PP, Union3, and DESY5, as indicated in the legend. Right panel: Same as the left panel, but for the \(w_{\dagger}\)VCDM model.}
    \label{PS_model_results}
\end{figure*}

Nonetheless, since the inferred values of \( H_0 \) from CMB only are in agreement with local measurements, we can reliably perform a joint analysis with the PP\&SH0ES dataset. The addition of this late-time data significantly tightens the constraints on all cosmological parameters. We find \( z_{\dagger} = 0.167^{+0.036}_{-0.045} \) and \( \Delta = -0.174\pm 0.054 \) at 68\% CL. These values indicate a departure from the \( \Lambda \)CDM dynamics, with a statistical preference for a late-time transition in the dark sector. The sign of \( \Delta < 0 \) implies that the dark energy equation of state transitions from a phantom-like regime at high redshifts to a quintessence regime at low redshifts. Moreover, the joint Planck+PP\&SH0ES dataset yields \( H_0 = 69.93 \pm 0.66 \) km/s/Mpc, effectively reducing the Hubble tension to a moderate level. For this joint analysis, we find a chi-square difference of \( \Delta \chi^2_{\text{min}} = -12.78 \), corresponding to a statistical significance of \( 3.1\sigma \), and an AIC difference of \( \Delta \text{AIC} = -8.78 \). These results collectively support the existence of a transition in the dark energy dynamics at low redshift. The interpretation of the results using the $w_{\dagger}$VCDM model is equivalent to that discussed previously.

Figure~\ref{PS_model_results} shows the marginalized one-dimensional posterior distributions and the 68\% and 95\% confidence level contours for the parameters $\Delta$, $H_0$, $S_8$, and $\Omega_m$ for both models considered in this work. The analysis incorporates data from Planck+PP\&SH0ES, Planck alone, and their combinations with DESI and several supernova samples, as indicated in the legend.

Now, let us focus on the joint analysis of Planck + DESI (see Table~\ref{tab:planck+bao}). We obtain \( \Delta = -0.226\pm 0.070 \) and \( z_{\dagger} = 0.533\pm 0.064 \), which represents a significant deviation from the \( \Lambda \)CDM model, favoring \( \Delta < 0 \) at the \( 3.2\sigma \) level. The values of \( z_{\dagger} \), as well as the overall parameter space of the model, are in tension with the Planck+PP\&SH0ES sample, as the latter uses the latest SH0ES Cepheid host distance anchors to calibrate the absolute magnitude of SN Ia. Given this tension, we opted not to combine PP\&SH0ES with DESI in our analysis.

It is also important to highlight that there is a noticeable change in the degeneracy direction in the $\Delta$–$H_0$ plane when the SH0ES Cepheid data are used, in contrast to other SN Ia samples. This can be interpreted as follows. It is well established that in models with standard $w$CDM dynamics—or in extended frameworks that effectively mimic such behavior—the joint analysis of CMB + SH0ES data tends to strongly prefer $w < -1$. In these cases, a clear negative correlation emerges in the $w$–$H_0$ plane (see, e.g., \cite{Escamilla:2023oce, Alestas:2020mvb, Silva:2025hxw, DiValentino:2019dzu, DiValentino:2016hlg}). In the context of the $w_{\dagger}$CDM and {$w_{\dag}$VCDM} models, the absence of any transition with $\Delta < 0$ corresponds to a pure phantom behavior.
However, even in the presence of a transition, the negative correlation in the $w$–$H_0$ plane is not expected to disappear for CMB + SH0ES, as the phantom regime may still persist over a significant redshift range, particularly for $z \gtrsim 0.5$ in the under consideration in this work. As a result, a joint analysis combining CMB and SH0ES data naturally favors values of $\Delta < 0$ to maintain a phantom behavior in that regime. This leads to an inversion in the degeneracy direction in parameter space—specifically in the correlations between $\Delta$, $H_0$, and $\Omega_m$—when compared to the results obtained using other SN Ia samples. This behavior is consistent with the statistical preference of the SH0ES dataset for higher $H_0$ values and its known negative correlation with $w$ in the $w$–$H_0$ plane. The inversion of the degeneracy direction in the $\Delta$–$H_0$ plane is therefore a direct reflection of this tension, emphasizing the different pulling effects that the SH0ES Cepheid calibration introduces in the global fit, especially when interpreted within a model that allows for evolving dark energy dynamics.

However, it is noteworthy that the overall dynamics favored by Planck+DESI align with those suggested by Planck+PP\&SH0ES. For the joint Planck + DESI analysis, we note that \( \Delta \text{AIC} = -10.18 \), again favoring our extended model. Considering that in our model the present-day equation of state \( w(z=0) \) plays the same role as the \( w_0 \) parameter in the CPL parametrization, we can directly compare our predictions with those from DESI-DR2~\cite{DESI:2025zgx}. In this case, we find \( w(z=0) = -0.774 \pm 0.070 \), which is in relative good agreement with the DESI-DR2 constraint \( w_0 = -0.42 \pm 0.21 \) with in 1.6$\sigma$.

We now turn to an analysis incorporating additional SN Ia datasets. Currently, three major SN Ia samples are available: PantheonPlus (PP), Union3, and DESY5. In Table~\ref{tab:planck+bao}, we summarize the results of combining each with Planck+DESI. These combinations consistently improve the parameter constraints and reinforce the statistical significance of a dynamical transition in the dark sector.

\begin{table*}[htpb!]
\begin{center}
\caption{Marginalized constraints, along with the mean values at 68\% CL, for both the free and some derived parameters of the models considered in this work, based on the CMB dataset and its combinations with DESI, PPS, PP, Union3, and DESY5. In the last rows, we present \(\Delta \chi^2_{\text{min}} = \chi^2_{\text{min (our model)}} - \chi^2_{\text{min ($\Lambda$CDM)}}\) and \(\Delta \text{AIC} = \text{AIC}_{\text{our model}} - \text{AIC}_{\text{$\Lambda$CDM}}\). Negative values for \(\Delta \chi^2_{\text{min}}\) and \(\Delta \text{AIC}\) indicate a better fit of the model compared to the \(\Lambda\)CDM model.}
\label{tab:planck+bao}
\renewcommand{\arraystretch}{1.2}
\resizebox{\textwidth}{!}{
\begin{tabular}{l||c|c|c|c} 
\hline
\textbf{Dataset} & \textbf{Planck+DR2} & \textbf{Planck+DR2+PP} & \textbf{Planck+DR2+Union3} & \textbf{Planck+DR2+DESY5} \\
\hline
\textbf{Model} & \textbf{$w_\dagger$CDM} & \textbf{$w_\dagger$CDM} & \textbf{$w_\dagger$CDM} & \textbf{$w_\dagger$CDM} \\
& \textcolor{blue}{\textbf{$w_{\dag}$VCDM}} & \textcolor{blue}{\textbf{$w_{\dag}$VCDM}} & \textcolor{blue}{\textbf{$w_{\dag}$VCDM}} & \textcolor{blue}{\textbf{$w_{\dag}$VCDM}} \\
\hline\hline
$10^{2}\omega{}_{b}$ & $2.241\pm 0.013$ & $2.249\pm 0.012$ & $2.245\pm 0.013$ & $2.245\pm 0.013$ \\
& \textcolor{blue}{$2.244\pm 0.013$} & \textcolor{blue}{$2.250\pm 0.012$} & \textcolor{blue}{$2.245\pm 0.012$} & \textcolor{blue}{$2.245\pm 0.012$} \\[0.1cm]

$\omega{}_{cdm}$ & $0.11929\pm 0.00070$ & $0.11836\pm 0.00058$ & $0.11893\pm 0.00060$ & $0.11892\pm 0.00056$ \\
& \textcolor{blue}{$0.11922^{+0.00079}_{-0.00070}$} & \textcolor{blue}{$0.11835\pm 0.00054$} & \textcolor{blue}{$0.11889\pm 0.00059$} & \textcolor{blue}{$0.11891\pm 0.00057$} \\[0.1cm]

$100\theta{}_{s}$ & $1.04198\pm 0.00025$ & $1.04206\pm 0.00028$ & $1.04201\pm 0.00025$ & $1.04197\pm 0.00028$ \\
& \textcolor{blue}{$1.04196\pm 0.00027$} & \textcolor{blue}{$1.04205\pm 0.00027$} & \textcolor{blue}{$1.04199\pm 0.00027$} & \textcolor{blue}{$1.04200\pm 0.00027$} \\[0.1cm]

$\ln10^{10}A_{s}$ & $3.049\pm 0.013$ & $3.052\pm 0.015$ & $3.048^{+0.014}_{-0.017}$ & $3.049\pm 0.015$ \\
& \textcolor{blue}{$3.045\pm 0.014$} & \textcolor{blue}{$3.050^{+0.013}_{-0.015}$} & \textcolor{blue}{$3.047\pm 0.014$} & \textcolor{blue}{$3.046^{+0.013}_{-0.015}$} \\[0.1cm]

$n_{s}$ & $0.9672\pm 0.0032$ & $0.9692\pm 0.0033$ & $0.9677\pm 0.0033$ & $0.9678\pm 0.0032$ \\
& \textcolor{blue}{$0.9675\pm 0.0034$} & \textcolor{blue}{$0.9693\pm 0.0032$} & \textcolor{blue}{$0.9680\pm 0.0033$} & \textcolor{blue}{$0.9682\pm 0.0033$} \\[0.1cm]

$\tau_{\text{reio}}$ & $0.0567^{+0.0059}_{-0.0068}$ & $0.0594^{+0.0067}_{-0.0077}$ & $0.0567^{+0.0065}_{-0.0079}$ & $0.0573\pm 0.0071$ \\
& \textcolor{blue}{$0.0555^{+0.0075}_{-0.0067}$} & \textcolor{blue}{$0.0584^{+0.0064}_{-0.0077}$} & \textcolor{blue}{$0.0566\pm 0.0069$} & \textcolor{blue}{$0.0561^{+0.0066}_{-0.0074}$} \\[0.1cm]

$\Delta$ & $-0.226\pm 0.070$ & $-0.102^{+0.031}_{-0.035}$ & $-0.174\pm 0.044$ & $-0.162\pm 0.031$ \\
& \textcolor{blue}{$-0.220\pm 0.077$} & \textcolor{blue}{$-0.103\pm 0.031$} & \textcolor{blue}{$-0.169\pm 0.043$} & \textcolor{blue}{$-0.161\pm 0.033$} \\[0.1cm]

$z_{\dagger}$ & $0.533\pm 0.064$ & $0.553^{+0.047}_{-0.140}$ & $0.524\pm 0.069$ & $0.490^{+0.063}_{-0.079}$ \\
& \textcolor{blue}{$0.521^{+0.057}_{-0.052}$} & \textcolor{blue}{$0.504\pm 0.095$} & \textcolor{blue}{$0.519\pm 0.067$} & \textcolor{blue}{$0.493^{+0.063}_{-0.081}$} \\

\hline
$\mathrm{H}_0 \, [\mathrm{km/s/Mpc}]$ & $65.3\pm 1.1$ & $67.13\pm 0.55$ & $66.06\pm 0.68$ & $66.40\pm 0.49$ \\
& \textcolor{blue}{$65.4\pm 1.1$} & \textcolor{blue}{$67.18\pm 0.53$} & \textcolor{blue}{$66.16\pm 0.68$} & \textcolor{blue}{$66.41\pm 0.50$} \\[0.1cm]

$\Omega_{\rm m}$ & $0.335\pm 0.012$ & $0.3141\pm 0.0058$ & $0.3256^{+0.0069}_{-0.0077}$ & $0.3222\pm 0.0051$ \\
& \textcolor{blue}{$0.333\pm 0.013$} & \textcolor{blue}{$0.3136\pm 0.0054$} & \textcolor{blue}{$0.3245\pm 0.0073$} & \textcolor{blue}{$0.3220\pm 0.0053$} \\[0.1cm]

$\mathrm{S}_{8}$ & $0.824\pm 0.011$ & $0.8077\pm 0.0083$ & $0.8163\pm 0.0090$ & $0.8154\pm 0.0079$ \\
& \textcolor{blue}{$0.821^{+0.012}_{-0.011}$} & \textcolor{blue}{$0.8068\pm 0.0080$} & \textcolor{blue}{$0.8153\pm 0.0084$} & \textcolor{blue}{$0.8145\pm 0.0079$} \\

\hline
$\Delta \chi^2_{\text{min}}$ & $-10.18$ & $-11.44$ & $-14.86$ & $-27.34$ \\
& \textcolor{blue}{$-12.40$} & \textcolor{blue}{$-11.40$} & \textcolor{blue}{$-16.36$} & \textcolor{blue}{$-29.24$} \\[0.1cm]

$\Delta{\text{AIC}}$ & $-6.18$ & $-7.44$ & $-10.86$ & $-23.34$ \\
& \textcolor{blue}{$-8.40$} & \textcolor{blue}{$-7.40$} & \textcolor{blue}{$-12.36$} & \textcolor{blue}{$-25.24$} \\

\hline
\hline
\end{tabular}}
\end{center}
\end{table*}

\begin{figure*}[htpb!]
    \centering
    \includegraphics[width=0.46\linewidth]{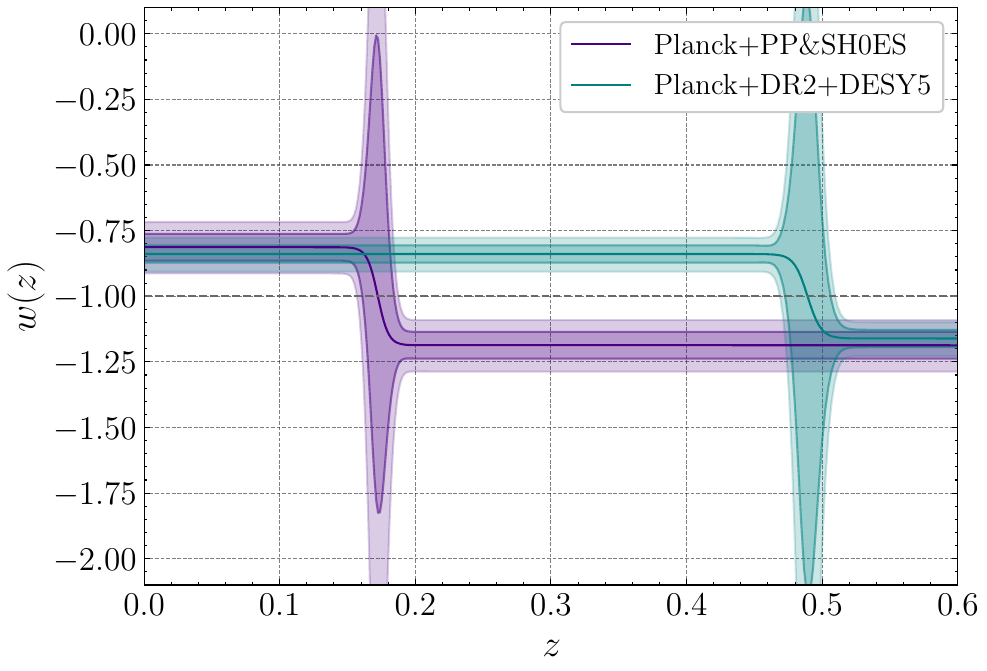}\,\,\,\,\,\,
    \includegraphics[width=0.46\linewidth]{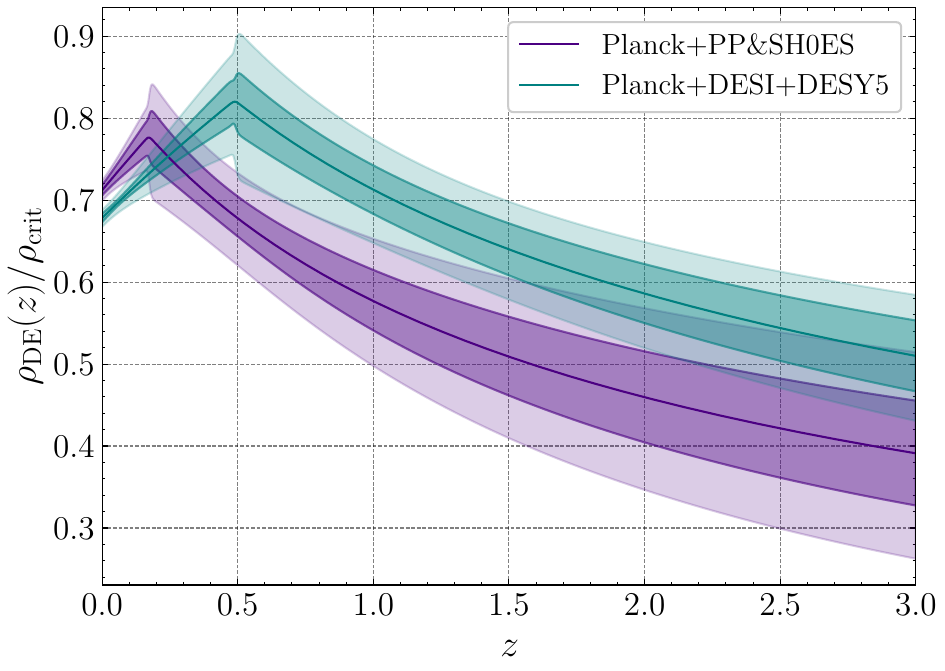}
        \caption{Left panel: Statistical reconstruction of the dark energy equation of state at 1$\sigma$ and 2$\sigma$ confidence levels based on the joint analyses of Planck + PP\&SH0ES and Planck + DESI + DESY5 data. The evolution shows a transition from a phantom-like regime ($w < -1$) at early times to a quintessence-like regime ($w > -1$) today, driven by the negative values of the deviation parameter $\Delta$. Right panel: Evolution of the normalized dark energy density, $\rho_{\text{DE}} / \rho_{\text{crit}}$, for the best-fit solutions from the same datasets. The plot highlights how the change in the deviation parameter affects the scaling of dark energy relative to the critical density at late times.}
    \label{w_and_rho_over_rhocrit_bestfit}
\end{figure*}

Let us first consider the combination Planck+DESI+PP. In this case, the inferred deviation parameter is \( \Delta = -0.102^{+0.031}_{-0.035} \) with a transition redshift of \( z_{\dagger} = 0.553^{+0.047}_{-0.140} \), showing a mild but statistically significant departure from \( \Lambda \)CDM at the \( 2.9\sigma \) level (\( \Delta \chi^2_{\text{min}} = -11.44 \), \( \Delta \text{AIC} = -7.44 \)). These results are consistent with the dynamic displayed by Planck+PP\&SH0ES, and strengthen the evidence for a late-time transition in the dark energy sector. The Hubble constant in this combination is constrained to \( H_0 = 67.13 \pm 0.55 \) km/s/Mpc, offering a partial alleviation of the \( H_0 \) tension, albeit not as strongly as with the inclusion of SH0ES calibration. In this case, our prediction for the present-day EoS is \( w(z=0) = -0.898^{+0.065}_{-0.031} \), which is in excellent agreement with the DESI-DR2 result, \( w_0 = -0.838 \pm 0.055 \) with in $0.71 \sigma$. 

Next, we analyze the Planck+DESI+Union3 combination. Here, we find \( \Delta = -0.174 \pm 0.044 \) and \( z_{\dagger} = 0.524 \pm 0.069\), favoring a dynamical transition at a significance level of \( 3.6\sigma \) (\( \Delta \chi^2_{\text{min}} = -14.86 \), \( \Delta \text{AIC} = -10.86 \)). Compared to the PP sample, Union3 supports a slightly more negative value of \( \Delta \), suggesting a more pronounced transition from phantom to quintessence behavior. The inferred Hubble constant is \( H_0 = 66.06 \pm 0.68 \) km/s/Mpc, consistent with intermediate redshift measurements and with slightly tighter constraints on \( \Omega_m = 0.3256^{+0.0069}_{-0.0077} \). These results further confirm the robustness of our model across multiple SN Ia datasets. The corresponding present-day EoS is \( w(z=0) = -0.826 \pm 0.044 \) at 68\% CL.

Finally, the most striking results are obtained when combining Planck+DESI+DESY5. In this case, we find the most significant statistical preference for our model: \( \Delta = -0.162 \pm 0.031 \) and \( z_{\dagger} = 0.490^{+0.063}_{-0.079} \), deviating from \( \Lambda \)CDM at the \( 4.9 \sigma \) (\(5.1\sigma\) level, with \( \Delta \chi^2_{\text{min}} = -27.34 \) \(-29.24\)) and \( \Delta \text{AIC} = -23.34 \). As highlighted in~\cite{Efstathiou:2024xcq}, a discrepancy of approximately 0.04 mag in magnitude between low and high redshifts was observed between the PP and DESY5 samples. This offset may partly explain the enhanced deviation seen in the DESY5 results. If the discrepancy is not due to systematic errors, it may point to genuine new physics in the dark sector, with our model capturing this late-time transition accurately. Lastly, the present-day EoS in this analysis predicts \( w(z=0) = -0.838 \pm 0.031 \) at 68\% CL.

To globally quantify the evolution of the dark energy equation of state (EoS), we present in the left panel of Figure~\ref{w_and_rho_over_rhocrit_bestfit} the statistical reconstruction of the EoS, along with the corresponding 1$\sigma$ and 2$\sigma$ confidence intervals, based on joint analyses of the Planck + PantheonPlus\&SH0ES and Planck + DESI DR2 + DES-Y5 datasets.
A clear phantom crossing is observed in the Planck + DESI + DES-Y5 combination, occurring around $z_{\dagger} \sim 0.5$, indicating a transition from a quintessence-like regime ($w > -1$) to a phantom regime ($w < -1$). This behavior is consistent with other joint analyses, with the exception of the Planck + PantheonPlus\&SH0ES combination, which—as previously discussed—shows significant tension with the others. In this particular case, the transition occurs earlier, at around $z_{\dagger} \sim 0.17$. These findings provide robust statistical evidence for a late-time transition in the nature of dark energy, from quintessence to phantom-like behavior. Such a transition is consistent with dynamical dark energy models and may have important implications for our understanding of the recent accelerated expansion of the Universe.

In the right panel of Figure~\ref{w_and_rho_over_rhocrit_bestfit}, we show the evolution of the normalized dark energy density, $\rho_{\rm DE} / \rho_{\rm crit}$, for the best-fit solutions obtained from the same datasets. The plot illustrates how variations in the deviation parameter affect the scaling behavior of dark energy relative to the critical density, particularly at late times. A pronounced peak appears around the transition redshift $z_{\dagger}$, marking the epoch at which the dark energy density reaches its maximum. For $z > z_{\dagger}$, as expected, the dark energy density decreases and asymptotically approaches zero, i.e., $\rho_{\rm DE} \rightarrow 0$. Crucially, none of our reconstructions exhibit any indication of $\rho_{\rm DE} < 0$ at any redshift, thereby allowing us to robustly rule out the possibility of a negative dark energy density.
For $z < z_{\dagger}$, the dark energy density evolves smoothly toward its present-day value, in agreement with the expectations from dynamical dark energy models.

As previously mentioned, all the analyses discussed above exhibit the same overall trends when interpreted within the $w_{\dagger}$VCDM framework. It is important to highlight that the $w_{\dagger}$VCDM model not only provides a slight improvement in the goodness of fit but also offers a solid theoretical foundation for the adopted parameterization. A major and noteworthy exception arises in the joint Planck+DESI+DESY5 analysis, where we observe an improvement at the level of approximately $5\sigma$ compared to the $\Lambda$CDM model. Such a high-significance result is particularly striking: in the context of statistical analyses, a $5\sigma$ detection is widely regarded as a robust threshold for a discovery-level claim. In Table~\ref{tab:chi-square}, we summarize the statistical significance of the results for both models across all analyses conducted in this study. Therefore, this deviation represents a substantial challenge to the dynamics predicted by the standard $\Lambda$CDM cosmological model, potentially signaling the need for extensions or modifications to the standard framework.

Figure~\ref{w_and_rho_over_rhocrit_bestfit} illustrates the evolution of the dark energy equation of state and the normalized energy density as functions of redshift. As seen in the left panel, the equation of state evolves from a phantom-like regime (\( w < -1 \)) at early times to a quintessence-like regime (\( w > -1 \)) today. This behavior is driven by the negative \( \Delta \) values inferred from all dataset combinations. Interestingly, this evolving behavior of the EoS is consistent with the trends reported in the latest DESI collaboration results \cite{DESI:2025zgx}, which also favor a dynamic dark energy scenario transitioning from phantom to quintessence. Moreover, the redshift at which this transition occurs in our model, lies within the same range indicated by DESI's findings—see Fig. 12 in Ref.~\cite{DESI:2025zgx}, around \( z \sim 0.3{-}0.5 \).
It is important to note that near the transition redshift, the rapid change predicted by the model leads to a significant amplification of error propagation within this interval. This effect does not impact the best-fit value itself but becomes evident when reconstructing the confidence regions, as it stems from the calculation of partial derivatives used in the statistical reconstruction of functions. In other words, this behavior arises purely from numerical effects associated with the sharp transition in the equation of state. The right panel complements this phenomenology by showing the impact of the transition on the background evolution, highlighting how the dark energy density tracks the critical density differently across epochs.


\begin{figure*}
    \centering
    \includegraphics[width=1\linewidth]{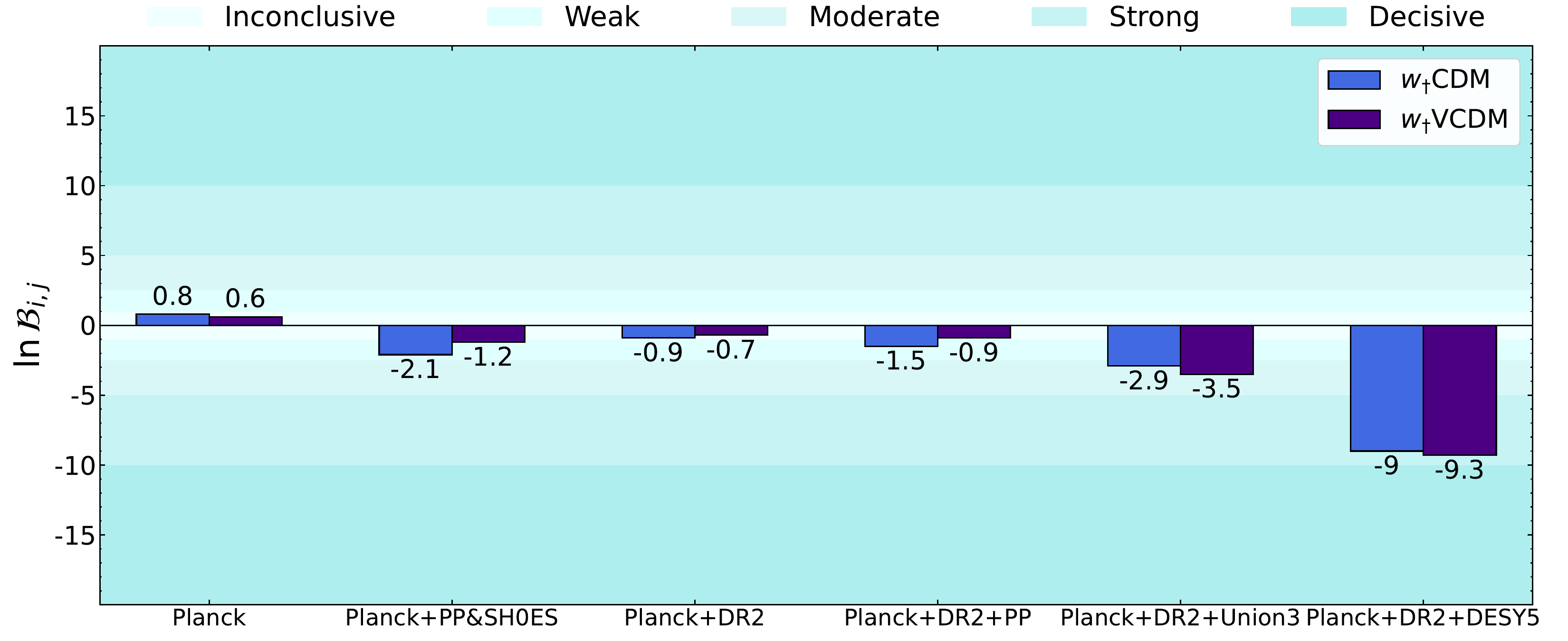}
    \caption{The values of $\ln \mathcal{B}_{ij}$ for all joint analyses performed in this work—covering both the $w_{\dagger}$CDM and $w_{\dagger}$VCDM models—are computed based on the statistical framework and methodology outlined in Section~\ref{Statistical_tests}.}
    \label{fig:bayes_factor}
\end{figure*}

Figure~\ref{fig:bayes_factor} displays the value of the logarithmic Bayes factor, $|\ln \mathcal{B}_{ij}|$, interpreted according to the Jeffreys–Kass–Raftery scale. This figure summarizes the strength of statistical evidence obtained for all analyses performed in this work, considering both phenomenological models under investigation. As expected, our most comprehensive joint analysis, Planck + DESI DR2 + DES-Y5, yields very strong evidence in favor of both the $w_{\dagger}$CDM and $w_{\dagger}$VCDM models when compared to the standard $\Lambda$CDM model. These findings underscore the robustness of the proposed frameworks, as supported by multiple statistical indicators, including goodness-of-fit criteria and Bayesian model comparison. The consistency of this evidence across independent datasets further reinforces the viability of late-time dynamical dark energy scenarios with a phantom transition around $z_{\dagger} \sim 0.5$, as encoded in our parametrizations and main results. As illustrated in the graph, additional secondary analyses consistently show strong and positive evidence.

\begin{figure*}[htpb!]
    \centering
    \includegraphics[width=0.46\textwidth]{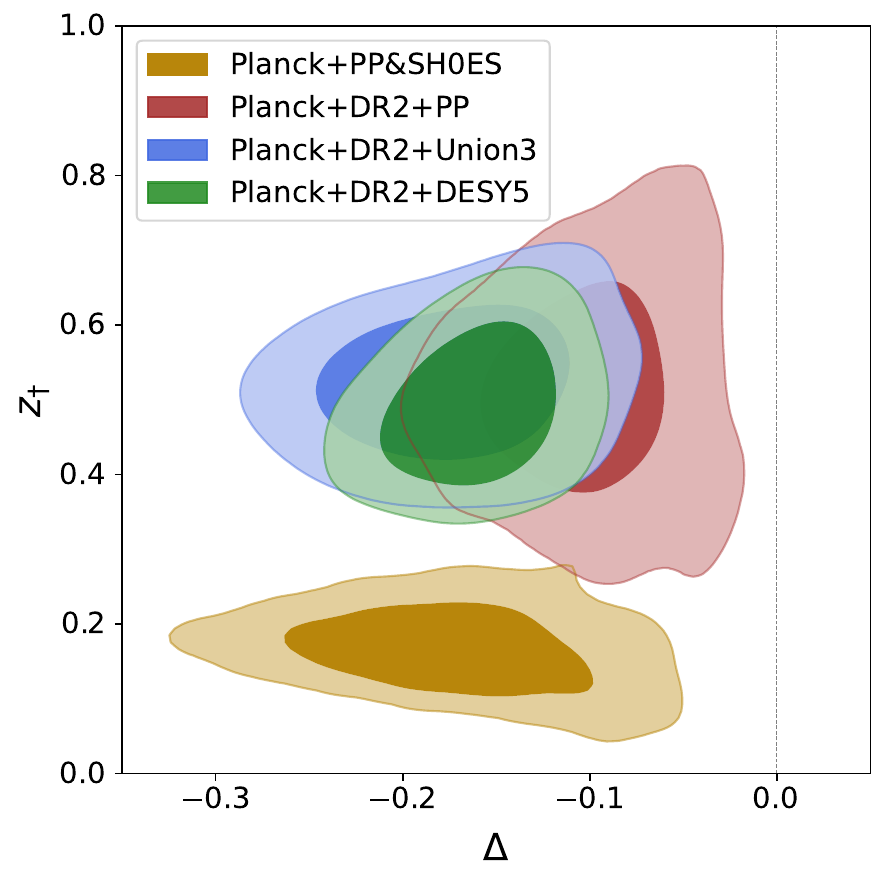}\,\,\,\,
    \includegraphics[width=0.46\textwidth]{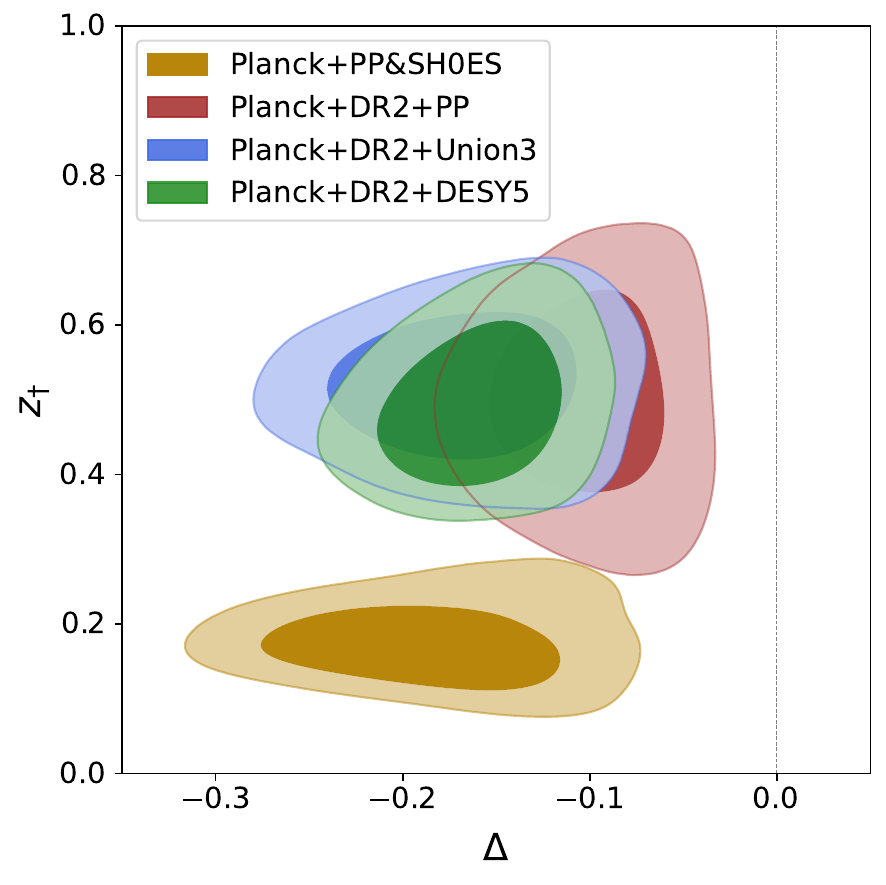}
    \caption{Left panel: The 2D contours at 68\% and 95\% confidence levels for the parameters \(\Delta\)–\(z_{\dagger}\) are obtained from different data combinations, as indicated in the legend, in the \(w_{\dagger}\)CDM model. Right panel: Same as the left panel, but for the \(w_{\dagger}\)VCDM model.}
    \label{PS_Delta_zt}
\end{figure*}


\begin{table}[htpb!]
\centering
\scalebox{0.9}{
\begin{tabularx}{1\linewidth}{l|c|c}
\hline

\textbf{$\bm{w_{\dagger}}$CDM} & \(\ \Delta \chi^2_{\text{min}}\) & \, Statistical significance $\sigma$ \\
\textcolor{blue}{\textbf{$\bm{w_{\dagger}}$VCDM}} & & \\
\hline
\hline
Planck & $-3.90$ & $1.5$\\
& \textcolor{blue}{$-3.62$} & \textcolor{blue}{$1.4$} \\[0.2cm]

Planck + DR2 & $-10.18$ & $2.7$ \\
& \textcolor{blue}{$-12.40$} & \textcolor{blue}{$3.1$}\\[0.2cm]

Planck + DR2 + PP & $-11.44$ & $2.9$ \\
& \textcolor{blue}{$-11.40$} & \textcolor{blue}{$2.9$} \\[0.2cm]

Planck + PP\&SH0ES & $-12.78$ & $3.1$\\
& \textcolor{blue}{$-13.20$} & \textcolor{blue}{$3.2$} \\[0.2cm]

Planck + DR2 + Union3 & $-14.86$ & $3.4$ \\
& \textcolor{blue}{$-16.36$} & \textcolor{blue}{$3.6$} \\[0.2cm]

Planck + DR2 + DESY5 & $-27.34$ & $4.9$ \\
& \textcolor{blue}{$-29.24$} & \textcolor{blue}{$5.1$} \\
\hline
\hline
\end{tabularx}
}
\caption{Statistical significance in favor of a transition from a phantom-like phase to a quintessence regime at late times, relative to the \(\Lambda\)CDM model.}
\label{tab:chi-square}

\end{table}

Although our model consistently yields a better fit to the data than the standard $\Lambda$CDM scenario—particularly in combinations involving DESY5 and PP\&SH0ES—it is important to emphasize that it neither resolves the $H_0$ tension nor substantially mitigates the $S_8$ discrepancy. As shown in Table~\ref{tab:chi-square}, the improvement in fit is statistically significant across all datasets, reaching the $\sim$5$\sigma$ level in the Planck+DESI+DESY5 analysis. Nevertheless, even in this best-fit case, the derived value of $H_0 = 66.40\pm 0.49$ km/s/Mpc remains in tension with local determinations of $H_0$, although it is in remarkable agreement with the DESI-DR2 prediction of \( H_{0} = 66.74 \pm 0.56 \) km/s/Mpc for the $w_{0}w_{a}$CDM model. This indicates that, while the model introduces a robust and theoretically well-motivated modification to the late-time behavior of dark energy—characterized by a transition from a phantom-like to a quintessence regime—it does not fully resolve all existing cosmological tensions.

\begin{figure*}[htpb!]
    \centering
    \includegraphics[width=0.46\textwidth]{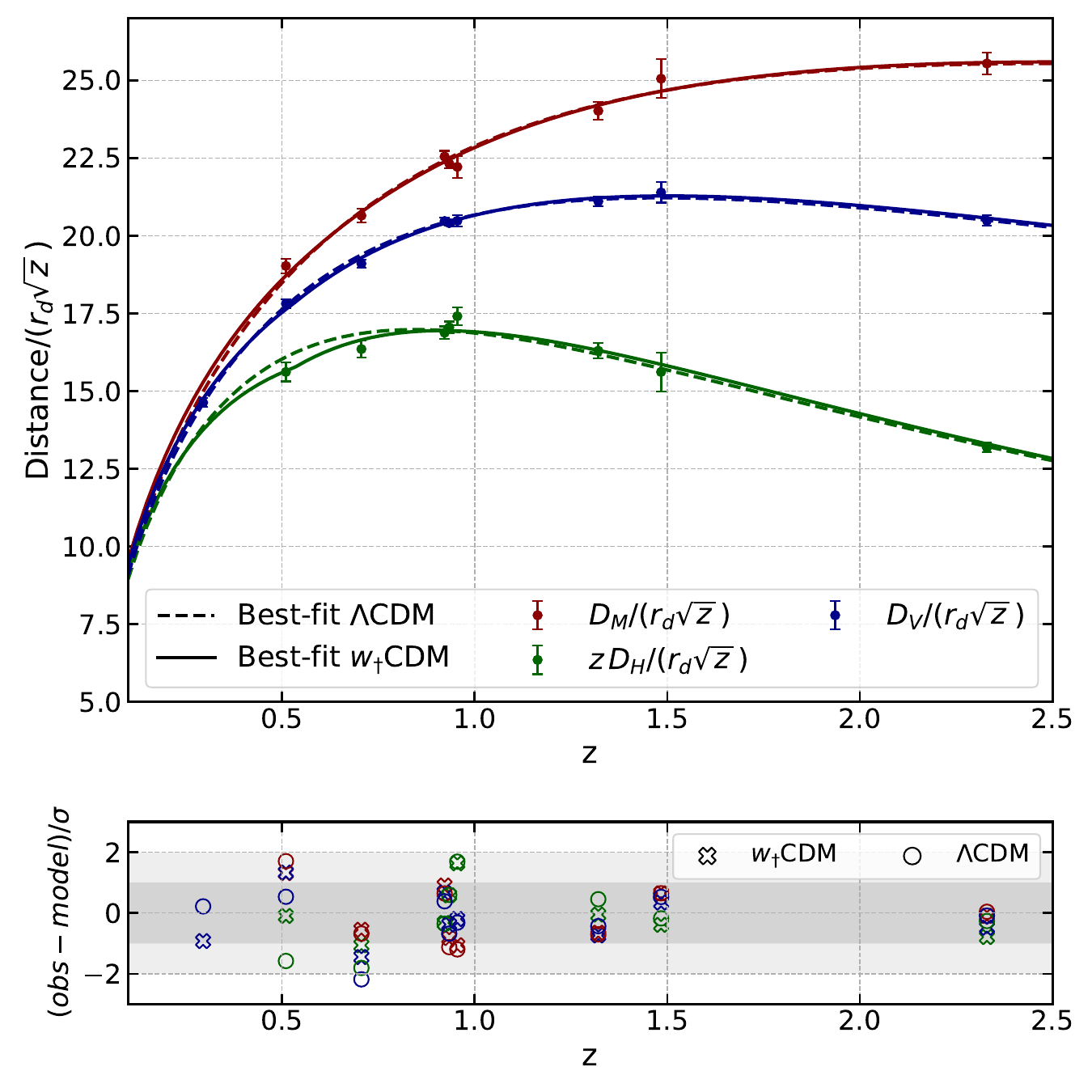}\,\,\,\,
    \includegraphics[width=0.46\textwidth]{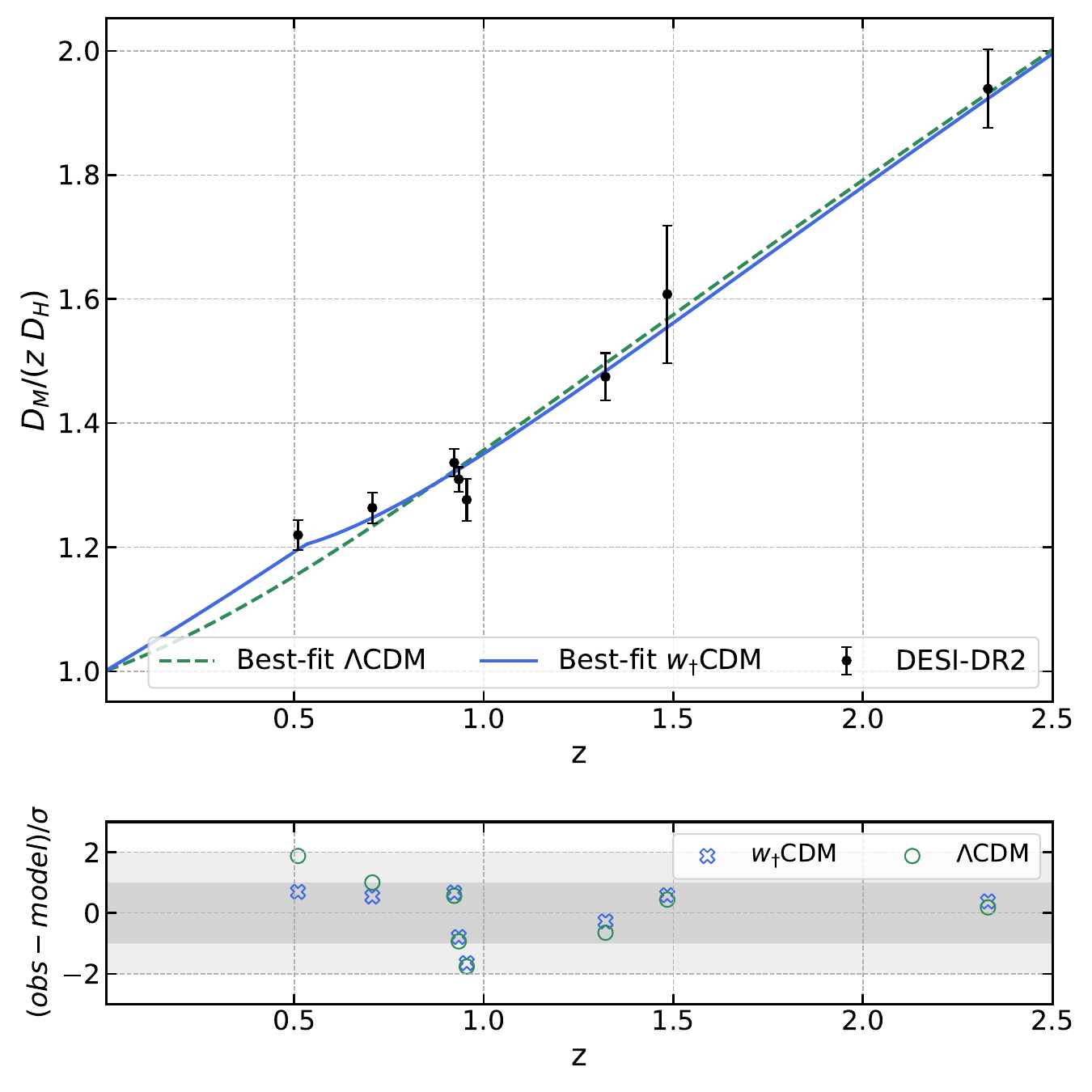}
    \caption{Best-fit curves for the rescaled distance–redshift relations in the $w_{\dagger}$CDM  (solid lines) and $\Lambda$CDM (dashed lines) models, based on the joint analysis of CMB and DESI DR2 datasets. The curves correspond to the three distance measures used in BAO observations, each distinguished by the colors indicated in the legend. Error bars reflect $\pm 1\sigma$ measurement uncertainties. Bottom panel: Deviations between the theoretical predictions and the observed BAO data, normalized by their respective observational errors. $w_{\dagger}$CDM results are marked with ‘x’ symbols, while $\Lambda$CDM results are marked with ‘o’ symbols.}
    \label{th_plot}
\end{figure*}

Figure \ref{th_plot} presents a comparison between the theoretical predictions of the models and the main observable considered in this work: the BAO measurements. The plot shows: (i) the best-fit predictions obtained from the joint analysis of CMB and DESI DR2 data for the main rescaled distance quantities probed by BAO; and (ii) the evolution of the ratio $D_M / (z D_H)$ as a function of redshift for both the $w_{\dagger}$CDM and $\Lambda$CDM models. The lower panel displays the normalized residuals between the DESI observational data and the model predictions—$\Lambda$CDM (shown as circles) and $w_{\dagger}$CDM (shown as crosses)—expressed in units of the observational uncertainty $\sigma$. Notably, for redshifts $z > 1$, both models exhibit nearly identical trends and provide comparable fits to the data. However, at lower redshifts, particularly for $z < 0.7$, their predictions begin to diverge. In this regime, the $w_{\dagger}$CDM model provides a significantly better fit to the measurements, especially around $z \sim 0.5$, where the standard $\Lambda$CDM model fails to capture the observed behavior. This theoretical trend highlights the ability of the proposed model to accurately capture the new BAO measurements at low redshifts.

\section{Conclusions}
\label{conclusions}

In this work, we have investigated an extension of the standard $\Lambda$CDM model that incorporates a late-time dynamical transition in the dark energy sector. This extended scenario introduces two additional parameters: the deviation parameter $\Delta$, which quantifies the departure from a cosmological constant behavior, and the transition redshift $z_\dagger$, marking the epoch at which the dark energy equation of state undergoes a transition—left unconstrained a priori and determined directly from the data. Through a comprehensive statistical analysis, we tested the observational viability of this model using multiple datasets, including CMB measurements from Planck, BAO data from DESI, and three independent SN Ia compilations (PantheonPlus, Union3, and DESY5), along with the locally calibrated PP\&SH0ES sample.

Our main findings can be summarized as follows:

\begin{itemize}
    \item Model Selection and Significance: The extended model consistently provides better fits than the standard $\Lambda$CDM scenario, exhibiting statistically significant improvements in both $\chi^2$ and AIC values. Across all dataset combinations, the $\Delta \chi^2$ values are negative and substantial, reaching as low as $-29.24$ for the Planck+DESI+DESY5 combination. Using a more robust metric known as the Bayes factor—which, in addition to accounting for the number of free parameters as in the AIC, also incorporates the full shape of the posterior—we find $\ln \mathcal{B}_{ij} \sim -9$ for both proposed models when considering the Planck+DESI+DES-Y5 dataset combination. This value corresponds to a statistical preference of approximately $5\sigma$ in favor of the extended models.

    \item Consistency Across Datasets: The values inferred for $\Delta$ and $z_\dagger$ are consistent across different dataset combinations, with all analyses favoring $\Delta < 0$ and $z_\dagger \sim 0.3{-}0.5$. These results suggest a coherent picture in which the DE component exhibits phantom-like behavior at high redshifts and transitions to a quintessence-like phase at low redshifts.

    \item Impact on Cosmological Tensions: The extended model results in only moderately higher values of $H_0$. For example, the Planck+PP\&SH0ES combination yields $H_0 = 69.93 \pm 0.66 $ km/s/Mpc, leading to a partial reduction of the Hubble tension but falling short of fully resolving it. Thus, while the model alleviates the discrepancy to some extent, it does not provide a satisfactory explanation for the $H_0$ tension. 

    \item Role of Supernova Samples (see Figure \ref{PS_Delta_zt}): Among the SN Ia datasets considered, the DESY5 sample provides the strongest statistical support for a dynamical transition in the dark sector. The enhanced deviation observed in this dataset, compared to PantheonPlus and Union3, suggests a possible signature of new physics. This strengthens the case for our proposed extension. Figure \ref{PS_Delta_zt} displays the 2D confidence contours at 68\% and 95\% CL in the $\Delta$–$z_\dagger$ plane, highlighting the distinct role played by the different supernova samples.
\end{itemize}

Our results open new avenues for exploring the nature of dark energy beyond the cosmological constant paradigm, as our analyses reveal deviations from the $\Lambda$CDM model at a significance level of approximately $5\sigma$. Notably, the inferred redshift range for the transition lies well within the observational capabilities of current surveys, such as DESI. While improved observational constraints will be crucial, future theoretical efforts should also focus on linking the phenomenological behavior uncovered in this work to specific microphysical mechanisms (see \cite{Wolf:2024stt,Wolf:2025jed,Ye:2024zpk,Ye:2025ulq,Smirnov:2025yru,Andriot:2025los,Hussain:2025vbo} for recent efforts in this direction). Embedding this effective description within a more fundamental theoretical framework would further strengthen its status as a physically motivated extension of the standard cosmological model.

\begin{acknowledgments}
\bigskip
\noindent We thank the referee for thoughtful feedback that contributed to improving both the clarity and significance of our results. M.S.\ and M.A.S received support from the CAPES scholarship. R.C.N. thanks the financial support from the Conselho Nacional de Desenvolvimento Científico e Tecnológico (CNPq, National Council for Scientific and Technological Development) under the project No. 304306/2022-3, and the Fundação de Amparo à Pesquisa do Estado do RS (FAPERGS, Research Support Foundation of the State of RS) for partial financial support under the project No. 23/2551-0000848-3.

\end{acknowledgments}

\bibliographystyle{apsrev4-1}
\bibliography{main}

\end{document}